\documentclass{mn2e}
\usepackage{graphicx}

\title{A {\it Chandra\/} observation of the interacting pair of
galaxies NGC 4485/NGC 4490}

\author[T.\,P. Roberts et al.]
{T.\,P. Roberts$^1$$^*$,
R.\,S. Warwick$^1$, M.\,J. Ward$^1$ \& S.\,S. Murray$^2$\\ $^1$X-ray
\& Observational Astronomy Group, Dept. of Physics \& Astronomy,
University of Leicester, University Road, Leicester, LE1 7RH\\
$^2$Harvard-Smithsonian Center for Astrophysics, 60 Garden Street,
Cambridge, MA 02138, USA\\
$^*$E-mail: tro@star.le.ac.uk}
\date{}

\def\ro{{\it ROSAT~\/}}
\def\asca{{\it ASCA~\/}}

\def\chan{{\it Chandra~\/}}

\def\ergcms{{\rm ~erg~cm^{-2}~s^{-1}}}

\def\ergsec{{\rm ~erg~s^{-1}}}

\def\atpcm{{\rm ~atoms~cm^{-2}}}
\def\ctsec{{\rm ~count~s^{-1}}}

\def\H0{{\rm ~km~s^{-1}~Mpc^{-1}}}

\def\eg{{\it e.g.~\/}}

\def\cf{{\it c.f.~\/}}
\def\la{\mathrel{\hbox{\rlap{\hbox{\lower4pt\hbox{$\sim$}}}{\raise2pt\hbox{$<$}}}}}
\def\ga{\mathrel{\hbox{\rlap{\hbox{\lower4pt\hbox{$\sim$}}}{\raise2pt\hbox{$>$}}}}}

\def\d25{D$_{25}$}
\def\nh{{$N_{H}$}}
\def\Ha{{H$\alpha~$}}
\def\hi{H {\small I}$~$}

\def\los{line-of-sight\thinspace}
\def\.25{0.25 keV\thinspace}
\def\lx{L$_{\rm X}$}

\begin{document}

\maketitle

\begin{abstract}
We report the results of a 20 ks \chan ACIS-S observation of the
galaxy pair NGC 4485/4490.  This is an interacting system containing a
late-type spiral with an enhanced star formation rate (NGC 4490), and
an irregular companion that possesses a disturbed morphology.  A total
of 29 discrete X-ray sources are found coincident with NGC 4490, but
only one is found within NGC 4485.  The sources range in observed
X-ray luminosity from $\sim 2 \times 10^{37}$ to $4 \times 10^{39}$
erg s$^{-1}$.  The more luminous sources appear, on average, to be
spectrally harder than the fainter sources, an effect which is
attributable to increased absorption in their spectra.  Extensive
diffuse X-ray emission is detected coincident with the disk of NGC
4490, and in the tidal tail of NGC 4485, which appears to be thermal
in nature and hence the signature of a hot ISM in both galaxies.
However, the diffuse component accounts for only $\sim 10\%$ of the
total X-ray luminosity of the system ($2 \times 10^{40}$ erg s$^{-1}$,
0.5 - 8 keV), which arises predominantly in a handful of the brightest
discrete sources.  This diffuse emission fraction is unusually low for
a galaxy pair that has many characteristics that would lead it to be
classified as a starburst system, possibly as a consequence of the
small gravitational potential well of the system.  The discrete source
population, on the other hand, is similar to that observed in other
starburst systems, possessing a flat luminosity function slope of
$\sim -0.6$ and a total of six ultraluminous X-ray sources (ULX).
Five of the ULX are identified as probable black hole X-ray binary
systems, and the sixth (which is coincident with a radio continuum
source) is identified as an X-ray luminous supernova remnant.  The ULX
all lie in star-formation regions, providing further evidence of the
link between the ULX phenomenon and active star formation.
Importantly, this shows that even in star forming regions, the ULX
population is dominated by accreting systems.  We discuss the
implications of this work for physical models of the nature of ULX,
and in particular how it argues against the intermediate-mass black
hole hypothesis.

\end{abstract}


\section{Introduction}

In the two years since its launch, the high spatial resolution ($\sim
0.5''$) X-ray optics of the \chan observatory have revolutionised the
study of spatially complex X-ray sources.  One particular area to
benefit greatly from this advance is the study of the X-ray properties
of nearby galaxies.  This is highlighted by the recent studies of
nearby starburst systems such as M82 (Matsumoto et al. 2001; Kaaret et
al. 2001), NGC 253 (Strickland et al. 2000) and the Antennae
(Fabbiano, Zezas \& Murray 2001), where populations of luminous
point-like sources and widespread, luminous diffuse emission
components are clearly resolved.  A further advantage of the high
spatial resolution offered by \chan is the sensitivity it provides to
faint point-like sources, reaching limiting fluxes at least an order
of magnitude fainter than in comparable exposures with previous
missions.  The result has been the detection of large numbers of X-ray
sources in nearby galaxies; for example 110 X-ray sources were
detected in the central regions of M101 to a limiting luminosity of
$\sim 10^{36} \ergsec$ (Pence et al. 2001), and 47 detected within the
bulge of NGC 1291 down to L$_{\rm X} \sim 10^{37} \ergsec$ (Irwin et
al. 2002).  This has, in turn, allowed the first detailed studies of
the luminosity functions of discrete X-ray source populations over a
wide range of galaxy types and luminosities.  For example, Tennant et
al. (2001) derive separate luminosity functions for the bulge and disk
regions of M81, showing that the disk luminosity function has a
shallower slope, consistent with a younger, more luminous source
population.  On a similar theme, Kilgard et al. (2002) compare the
luminosity functions of starburst, disk-dominated and bulge-dominated
spiral galaxies, showing that galaxies hosting enhanced star formation
have flatter luminosity function slopes due, primarily, to the
presence of younger, more luminous X-ray sources in starburst systems.

In this paper we focus upon a new \chan observation of the nearby ($d
= 7.8$ Mpc, Tully 1988\footnote{We note that Tully gives separate
distances of 7.8 and 9.3 Mpc for NGC 4485 and NGC 4490 respectively.
This is obviously unphysical for a closely interacting system, so we
assume a distance of 7.8 Mpc for both galaxies throughout this paper,
as taken by Elmegreen et al. (1998).  A consistent distance of 8 Mpc
is assumed by Viallefond, Allen \& de Boer (1980), based on the
suspected membership of NGC 4485/90 to the CVn {\small II} cloud (de
Vaucouleurs, de Vaucouleurs \& Corwin 1976).}) interacting galaxy pair
NGC 4485/NGC 4490.  NGC 4490 is a late-type spiral galaxy, classified
as type SB(s)d with absolute magnitude $M_B = -19.55$ (de Vaucouleurs
et al. 1991).  Its smaller companion NGC 4485 is an irregular galaxy
(type IB(s)m) with $M_B = -17.65$.  At the quoted distance, their
projected separation is a mere $\sim 8$ kpc.  The closest encounter
between the systems occurred about $4 \times 10^8$ years ago (based on
a comparison with $N$-body simulations, and the age of the stellar
populations; Elmegreen et al. 1998), and was a prograde encounter that
has produced tidal features in both galaxies, including a ``bridge''
of material linking a tidal tail at the south of NGC 4485 to the
western arm of NGC 4490, and a faint tidal tail to the east of NGC
4490.

NGC 4490 shows considerable evidence for enhanced star formation in
both radio continuum and far infra-red observations (Viallefond, Allen
\& de Boer 1980; Klein 1983; Thronson et al. 1989).  Clemens,
Alexander \& Green (1999) suggest that it is a relatively young
galaxy, formed $2 \times 10^9$ years ago, with an approximately
constant star formation rate of $\sim 5$ M$_{\odot}$yr$^{-1}$
throughout its lifetime.  This activity predates the interaction with
NGC 4485, which has only had time to significantly affect the western
arm of NGC 4490.  The star formation appears to have driven a
galactic-scale bi-polar outflow of \hi gas perpendicular to the plane
of NGC 4490, resulting in a \hi envelope $\sim 56$ kpc across ($\equiv
25'$ on the sky) surrounding the galaxies (Clemens, Alexander \& Green
1998).  The passage of NGC 4485 through this \hi cloud appears to be
ram-pressure stripping it of its atomic, molecular and dusty ISM
components (Clemens, Alexander \& Green 2000).  Detailed \Ha imaging
(Duval 1981; Thronson et al. 1989) shows the ongoing star formation in
the system to be predominantly in the tidal arms between NGC 4485 and
NGC 4490, with additional activity peaking slightly to the west of the
nucleus of NGC 4490, and in a northern spiral arm.

This pair of galaxies is not particularly well studied in X-rays.  The
most detailed study to date was undertaken as part of a survey of
nearby spiral galaxies with the \ro PSPC by Read, Ponman \& Strickland
(1997; hereafter RPS97).  They detect four compact X-ray sources
embedded in extensive diffuse emission.  Three sources are associated
with the disk of NGC 4490, including one coincident with the nucleus,
and the fourth is located at the southern tip of NGC 4485.  The
spectrum of the diffuse emission appears unusually hard, leading to
the speculation that it is largely composed of unresolved X-ray binary
systems.  Subsequently, two \ro HRI images of the galaxies were
analysed by Roberts \& Warwick (2000; hereafter RW2000) as part of a
survey of point-like X-ray sources in nearby galaxies.  They detected
the same four sources in each observation, plus a fifth apparently
transient source on the southern side of NGC 4490, which is only
present in the first HRI observation.

Here, we present the results of a recent \chan ACIS-S observation of
the NGC 4485/90 system.  We sub-divide the paper in the following
manner.  In Section 2 we give details of the observation and data
reduction.  We catalogue the luminous discrete X-ray source population
of the galaxies in Section 3, and then investigate the X-ray
properties of the population in Section 4, which includes an
examination of the properties of six ``ultraluminous X-ray sources''
(ULX) found within the system.  This is followed by the analysis of
the remaining diffuse X-ray emission in Section 5.  In Section 6 we
discuss the total luminosity of the galaxies and the luminosity
function of the constituent sources, and compare our results to other
starburst systems observed by {\it Chandra\/}.  We go on to discuss
the implications of the discovery of so many ULX in this particular
small galaxy pair for our understanding of the ULX phenomenon in
Section 7 before, finally, presenting our conclusions.

\section{The Chandra observation}

The NGC 4485/90 system was observed by \chan on 2000 November 3 in a
single exposure with a total integration time of 19522 seconds.  The
nominal pointing position was at $12^h30^m31.2^s, +41^{\circ}39'00''$,
west of the nucleus of NGC 4490.  Both galaxies were positioned such
that virtually all their optical extent (defined by the D$_{25}$
elliptical isophote), excepting small regions at the extreme
north-east of NGC 4485 and east of NGC 4490, was encompassed within
the area covered by the back-illuminated S3 chip.  The ACIS-S
instrument was operated in the standard full-frame mode, with an
integration time of 3.2 s; recorded events were telemetered in the
faint mode.

Data reduction was performed using the {\small CIAO} software, version
2.1, starting with the level 2 event files.  These were filtered such
that all events with energies outside the 0.3 -- 10 keV range were
rejected.  The in-orbit background was at a constant low-level during
the observation, so that no filtering of periods of background flare
contamination was required.  Further steps in the analysis of the data
are outlined in the appropriate sections below\footnote{Note that many
of the analysis methods used in this paper follow the standard \chan
data analysis threads, published by the \chan X-ray Center on {\tt
http://asc.harvard.edu/ciao/}.}.

\section{The discrete source population}

A 0.3 -- 10 keV \chan image of the NGC 4485/90 system was produced
from the cleaned events file using the standard {\small CIAO} tools.
The image was created at the full spatial resolution of 0.492
arcsecond/pixel over a $1000 \times 1000$ pixel ($\equiv 8.2 \times
8.2$ arcminute) grid centred at the raw pixel coordinate $x = 4100, y
= 4100$ ($\equiv 12^h30^m34.2^s, +41^{\circ}39'16''$).  Further images
were similarly produced in the 0.3 -- 2 keV (hereafter ``soft'') and
the 2 -- 10 keV (``hard'') energy bands.

The detection of discrete X-ray sources was performed using the
{\small WAVDETECT} package.  This was run over the 2, 4, 8 and 16
pixel wavelet scales to a significance threshold of $1 \times
10^{-6}$, equivalent to an average of one false detection over our
field of view.  This range of wavelet scales allows the detection of
both point-like and moderately-extended sources ($\sim 10''$ diameter
on-axis).  {\small WAVDETECT} was initally run on the full band image,
producing 42 detections.  After a visual examination of the
detections, we limited our source list to a detection significance of
3.5$\sigma$ or better, leaving the 36 most significant detections.  We
also ran {\small WAVDETECT} on the soft and then the hard band images
individually, detecting 33 and 22 of the full band sources
respectively to the same detection criteria (20 of which were detected
in both bands).  There were no examples of a source detected
exclusively in either the soft or hard band and not in the full band;
this is a direct consequence of both the excellent spatial resolution
of \chan and the moderate exposure of the image, which results in a
background per detection element of close to zero for all bands.  One
source (CXOU J123024.0+413840) was detected only in the full band
image, being just below the detection threshold in both the soft and
hard images.

\begin{table*}
\caption{The 36 detected X-ray sources in the central region of the
NGC 4485/90 \chan field.}
\begin{tabular}{lcccccl}\hline
Source designation	& \multicolumn{2}{c}{Full band}
& Soft band	& Hard band	& Hardness ratio	& Other IDs/comments \\
CXOU J	& Count rate	& Detection	& \multicolumn{2}{c}{Count
rate}	& (HR) \\
	& ($\times 10^{-3} \ctsec$)	& signif. ($\sigma$)
& \multicolumn{2}{c}{($\times 10^{-3} \ctsec$)}	&	& \\\hline
123014.8+414145  &   0.7 $\pm$ 0.2 &   6.1 & 0.4 $\pm$ 0.2 &  [0.3 $\pm$ 0.1] & -0.21 $\pm$ 0.27 &Background source \\
123015.2+413535  &   0.6 $\pm$ 0.2 &   5.1 & 0.6 $\pm$ 0.2 &  [0.1 $\pm$ 0.1] & -0.62 $\pm$ 0.23 &Background source \\
123017.6+413849  &   0.5 $\pm$ 0.2 &   4.5 & 0.2 $\pm$ 0.1 &  0.3 $\pm$ 0.1 &  0.24 $\pm$ 0.31 &Background source\\
123023.5+413652  &   3.0 $\pm$ 0.4 &  22.0 & 2.1 $\pm$ 0.3 &  0.9 $\pm$ 0.2 & -0.42 $\pm$ 0.12 &Background source \\
123024.0+413840  &   0.4 $\pm$ 0.2 &   3.5 & [0.2 $\pm$ 0.1] &  [0.2 $\pm$ 0.1] & -0.08 $\pm$ 0.38 & \\
123025.3+413924  &   3.3 $\pm$ 0.5 &  25.4 & 2.2 $\pm$ 0.4 &  1.1 $\pm$ 0.3 & -0.32 $\pm$ 0.12 & \\
123027.1+413929  &   0.5 $\pm$ 0.2 &   3.9 & 0.3 $\pm$ 0.1 &  [0.2 $\pm$ 0.1] & -0.30 $\pm$ 0.35 & \\
123027.3+413813  &   1.0 $\pm$ 0.2 &   8.1 & 0.2 $\pm$ 0.1 &  0.8 $\pm$ 0.2 &  0.55 $\pm$ 0.19 & \\
123028.3+413958  &   1.1 $\pm$ 0.3 &   7.6 & 1.1 $\pm$ 0.3 &  [0.1 $\pm$ 0.1] & -0.86 $\pm$ 0.12 & \\
123028.7+413926  &   0.8 $\pm$ 0.2 &   5.6 & 0.3 $\pm$ 0.1 &  0.5 $\pm$ 0.2 &  0.31 $\pm$ 0.26 & \\
123028.8+413756  &   0.6 $\pm$ 0.2 &   5.0 & 0.5 $\pm$ 0.2 &  [0.1 $\pm$ 0.1] & -0.56 $\pm$ 0.25 & \\
123029.2+414046  &   0.9 $\pm$ 0.2 &   7.2 & 0.7 $\pm$ 0.2 &  [0.3 $\pm$ 0.1] & -0.43 $\pm$ 0.21 & Bridge source\\
123029.5+413927  &  14.6 $\pm$ 1.1 &  83.3 & 6.5 $\pm$ 0.7 &  8.1
$\pm$ 0.8 &  0.11 $\pm$ 0.06 & FIRST J123029.4+413927 $^a~^b$ \\
123030.4+413852  &   8.2 $\pm$ 0.8 &  44.4 & 4.8 $\pm$ 0.6 &  3.5 $\pm$ 0.5 & -0.16 $\pm$ 0.08 & \\
123030.6+414142  &  76.1 $\pm$ 4.3 & 336.5 &50.9 $\pm$ 3.0 & 25.1
$\pm$ 1.7 & -0.34 $\pm$ 0.02 & NGC 4485 X-1 $^b~^c$\\
123030.8+413911  &  36.5 $\pm$ 2.3 & 167.0 &19.1 $\pm$ 1.4 & 17.4
$\pm$ 1.3 & -0.05 $\pm$ 0.04 & Transient $^b$\\
123031.1+413837  &   0.8 $\pm$ 0.2 &   5.2 & [0.4 $\pm$ 0.2] &  0.4 $\pm$ 0.1 & -0.06 $\pm$ 0.29 & \\
123031.4+413901  &   0.6 $\pm$ 0.2 &   3.6 & [0.3 $\pm$ 0.2] &  0.4 $\pm$ 0.1 &  0.13 $\pm$ 0.34 & \\
123032.3+413918  &  27.5 $\pm$ 1.8 & 136.3 &17.1 $\pm$ 1.3 & 10.3
$\pm$ 0.9 & -0.25 $\pm$ 0.04 & NGC 4490 X-1 $^b~^c$\\
123034.3+413805  &   1.3 $\pm$ 0.3 &   9.5 & 0.6 $\pm$ 0.2 &  0.6 $\pm$ 0.2 &  0.04 $\pm$ 0.21 & \\
123034.4+413849  &   6.9 $\pm$ 0.7 &  41.2 & 5.1 $\pm$ 0.6 &  1.8 $\pm$ 0.3 & -0.47 $\pm$ 0.08 & \\
123035.2+413846  &   2.0 $\pm$ 0.4 &  11.7 & 1.4 $\pm$ 0.3 &  0.5 $\pm$ 0.2 & -0.50 $\pm$ 0.15 & \\
123035.9+413832  &   0.9 $\pm$ 0.2 &   4.8 & 0.6 $\pm$ 0.2 &  [0.2 $\pm$ 0.1] & -0.60 $\pm$ 0.25 & \\
123036.3+413837  &  29.2 $\pm$ 1.9 & 131.5 &18.3 $\pm$ 1.3 & 10.9
$\pm$ 0.9 & -0.25 $\pm$ 0.04 & NGC 4490 X-2 $^b~^c$\\
123038.3+413830  &   0.5 $\pm$ 0.2 &   3.8 & 0.6 $\pm$ 0.2 &  [0.0 $\pm$ 0.0] & -1.00 $\pm$ 0.16 & \\
123038.5+413742  &   2.8 $\pm$ 0.4 &  19.9 & 2.1 $\pm$ 0.4 &  0.6
$\pm$ 0.2 & -0.55 $\pm$ 0.12 & NGC 4490 X-3 $^c$\\
123039.0+413809  &   2.7 $\pm$ 0.4 &  19.8 & 1.7 $\pm$ 0.3 &  1.0 $\pm$ 0.2 & -0.23 $\pm$ 0.14 & \\
123039.1+413822  &   0.7 $\pm$ 0.2 &   4.6 & 0.5 $\pm$ 0.2 &  [0.1 $\pm$ 0.1] & -0.78 $\pm$ 0.29 & \\
123039.1+413751  &   0.5 $\pm$ 0.2 &   3.8 & 0.5 $\pm$ 0.2 &  [0.0 $\pm$ 0.0] & -1.00 $\pm$ 0.15 & SSS\\
123040.4+413813  &   5.9 $\pm$ 0.6 &  34.7 & 3.7 $\pm$ 0.5 &  2.1 $\pm$ 0.4 & -0.26 $\pm$ 0.09 & \\
123043.1+413756  &   0.8 $\pm$ 0.2 &   6.0 & 0.8 $\pm$ 0.2 &  [0.0 $\pm$ 0.1] & -0.96 $\pm$ 0.14 & \\
123043.2+413818  &  50.7 $\pm$ 3.0 & 259.4 &29.8 $\pm$ 1.9 & 20.7
$\pm$ 1.5 & -0.18 $\pm$ 0.03 & NGC 4490 X-4 $^b~^c$\\
123045.6+413639  &   1.1 $\pm$ 0.2 &   8.4 & 0.9 $\pm$ 0.2 &  [0.1 $\pm$ 0.1] & -0.75 $\pm$ 0.16 & \\
123047.8+413807  &   0.5 $\pm$ 0.2 &   4.2 & 0.3 $\pm$ 0.1 &  [0.2 $\pm$ 0.2] & -0.18 $\pm$ 0.42 & \\
123047.9+413727  &   2.0 $\pm$ 0.3 &  15.4 & 1.6 $\pm$ 0.3 &  0.4 $\pm$ 0.1 & -0.63 $\pm$ 0.13 & \\
123049.6+414056  &  12.1 $\pm$ 1.0 &  89.2 & 9.5 $\pm$ 0.8 &  2.6 $\pm$ 0.4 & -0.57 $\pm$ 0.05 &
Background source (S2 chip)\\\hline
\end{tabular}
\begin{tabular}{l}
Notes: $^a$ Probable supernova remnant.  $^b$ ULX.  $^c$ Previous
X-ray source ID from RW2000.\\
\end{tabular}
\label{srcs}
\end{table*}

The sources are listed in Table~\ref{srcs}.  In this table we show
their name according to the official IAU designation for \chan sources
(which incorporates their position in J2000 coordinates), the count
rates in the full, soft and hard bands, the significance of their full
band detection and a hardness ratio (discussed below).  Where {\small
WAVDETECT} did not detect a source in a particular band, we quote a
count rate based on aperture photometry at the position of the source,
using a 4-pixel radius to extract the source counts and an annulus
between 4 and 8 pixels in radius to provide a measure of the local
background for each source.  These aperture-based count rates are
shown in Table~\ref{srcs} in square brackets\footnote{The
aperture-based count rates are not corrected for systematic errors
such as the encircled energy fraction.  This is because for these low
count rates ($\leq 6$ counts in 19522 s) and small off-axis angles the
corrections are significantly smaller than the error on the count
rates.}.  We also add comments on possible identifications based on
previous observations, multi-wavelength data and the source position
on the sky.

We assume a flat exposure time of 19522s across the field.  This is
because the exposure map calculated for the observation at 1.5 keV
shows that, over the area on the S3 chip covered by the galaxies, the
effective exposure varies by no more than 5\%.  To correct for
exposure variations one would ideally create a single, multi-energy
exposure map appropriate to all sources within the galaxies.  However,
this is not sensible, since the sources vary greatly in spectral
shape, whilst a multi-energy exposure map would of necessity
incorporate an ``average'' spectral form.  Instead, we simply add the
5\% error, in quadrature, to the existing counting errors.  This
systematic error only exceeds the intrinsic error in the five
brightest detected sources, and is negligible for the faint sources.

The sources were all examined for spatial extension beyond the \chan
point spread function.  However, under close scrutiny an analysis of
the source extent was found to be suitable only in the case of the
brightest sources, where the photon statistics were sufficient for a
detailed study.  We present these results in \S 4.2.1.

\begin{figure*}
\centering
\includegraphics[width=12cm,angle=270]{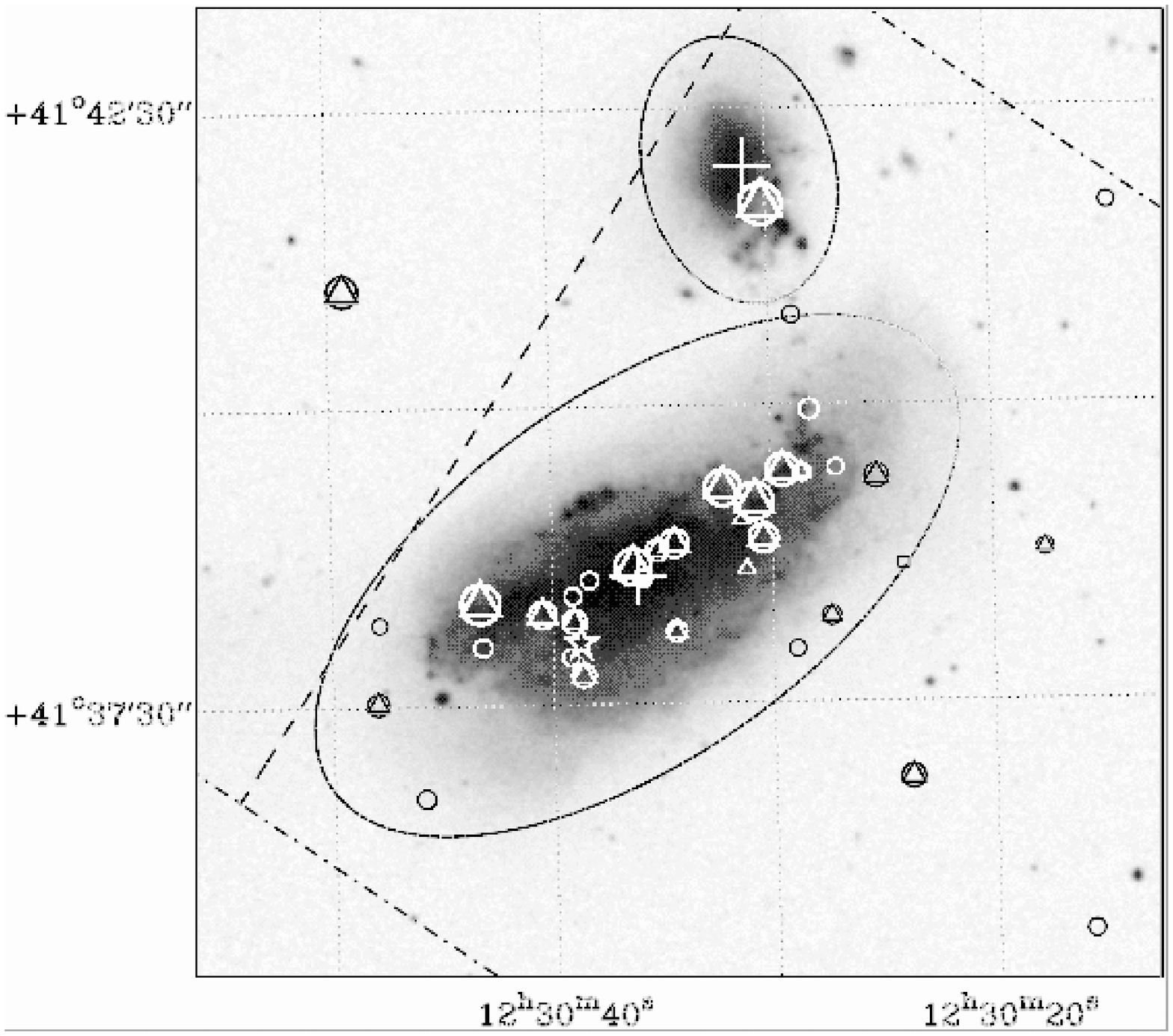}
\caption{The \chan X-ray source detections overlayed
on the DSS-2 blue image of NGC 4485/4490.  Soft band detections are
marked by open circles, and hard band detections by open triangles.
The size of the markers scales logarithmically with the source count
rates.  CXOU J123024.0+413840 is shown by an open square.  The optical
nucleus of each galaxy is denoted by a large cross, and the position
of SN 1982F by an open star.  The extent of the \d25 ellipses, the
limits of the ACIS-S array (dot-dash lines) and the divide between the
S2 and S3 chips (dashed line, with S2 to the top left) are also
shown.}
\label{dss2srcs}
\end{figure*}

Figure~\ref{dss2srcs} shows the positions of the X-ray sources
overlayed on to a digitised Palomar sky survey image (taken from the
DSS-2 blue data), with hard and soft band detections differentiated by
the use of separate symbols.  This figure also shows the optical
extent of the galaxies (shown by the \d25 ellipse, representing the 25
mag/arcsecond$^2$ isophote for each galaxy) and the regions covered by
the ACIS-S S2 and S3 chips.  Note that the size of the markers does
not in any way represent the positional uncertainty of the sources;
the ACIS-S positions are accurate to an RMS of $< 0.6''$ (\chan
Proposer's Observatory Guide), which is a factor 10 smaller than the
smallest symbol in the figure.

We see a total of 29 sources within the \d25 ellipse of NGC 4490, with
only one source (albeit the brightest), CXOU J123030.6+414142,
coincident with NGC 4485.  Intriguingly, a further source - CXOU
J123029.2+414046 - is located between the galaxies (as demarkated by
their \d25 ellipses), and may be associated with a bridge of material
pulled out of NGC 4485 by the tidal interaction, as described by
Elmegreen et al. (1998).  There are a total of five probable
background sources, the brightest of which lies on the S2 chip.  If we
consider the photometric nuclear positions of both galaxies listed in
NED (from Falco et al. 1999), no evidence is seen for an X-ray source
coincident with the either nucleus; in the case of NGC 4490 two
sources are detected in the proximity of the nucleus, but are
separated from the nuclear position by 6$'' (\equiv 220$pc; CXOU
J123035.9+413832) and 9$'' (\equiv 340$pc; CXOU J123036.3+413837)
respectively (c.f. an error of 2.5$''$ on the photometric nuclear
positions).  We also do not detect an X-ray counterpart to the recent
supernova SN 1982F, though there are several X-ray sources in its
close proximity.

Of the 30 sources coincident with the galaxies, we would expect some
fraction to be background objects.  We investigate the level of this
contamination using the \chan Deep Field South number counts of
Giacconi et al. (2001), treating the soft and hard bands separately.
In the hard (2 -- 10 keV) band, Giacconi et al. (2001) measure an
average spectral index for cosmic X-ray sources of $\Gamma = 1.35$.
Using the same spectral form, we obtain a limiting flux of $7.8 \times
10^{-15} \ergcms$ for our limiting count rate of $3 \times 10^{-4}
\ctsec$.  This converts to a total of 308 sources per square degree
using Equation (2) of Giacconi et al. (2001), or a total contamination
of $< 1.6$ cosmic background sources over the combined area of NGC
4485 and NGC 4490.  This amounts to fewer than 10\% of the 19 hard
band detections.

The situation in the soft band is somewhat complicated by the
intrinsic \hi distribution of the galaxy pair (e.g. Clemens, Alexander
\& Green 1998).  The minimum line-of-sight \hi column through the
galaxies at the \d25 isophote is $6 \times 10^{20} \atpcm$, which
increases substantially towards the centre of NGC 4490.  Hence, the
flux detection limit for any cosmic sources lying behind the galaxy
will be severely affected.  We account for this by converting our
minimum soft band (0.3 -- 2 keV) count rate into an unabsorbed 0.5 --
2 keV flux (assuming $\Gamma = 1.26$ from the full band powerlaw fit
of Giacconi et al. (2001), and a line-of-sight column composed of
their best-fit value plus a typical (minimum) H~{\small I} column for
the source detections of $1.4 \times 10^{21} \atpcm$).  This gave an
equivalent cosmic source flux detection limit of $6.7 \times 10^{-16}
\ergcms$.  Equation (1) of Giacconi et al. then implies that we have a
cosmic source contamination of $< 4.8$ sources in the soft band, which
is almost certainly an overestimate due to the fact that the \hi and
X-ray source distributions appear strongly correlated.  In comparison,
there are 28 soft band source detections coincident with the galaxies.

The above numbers can be confirmed using the off-galaxy regions
covered by the S3 chip in our \chan observation.  In total, an area
1.6 times greater than that of the galaxies is covered off-galaxy on
the S3 chip (see Figure~\ref{dss2srcs}), at a similar flux limit.  A
total of 5 soft and 2 hard cosmic sources are found within this
region, consistent with the above contamination estimates.  We
therefore consider that the total contamination of the 30 NGC 4485/90
X-ray sources by a cosmic component to be at most $\sim 15\%$, and
probably less.

\section{Characteristics of the discrete X-ray sources}

\subsection{Hardness ratios and spectral trends in NGC 4490}

\begin{figure}
\centering
\includegraphics[width=6cm,angle=270]{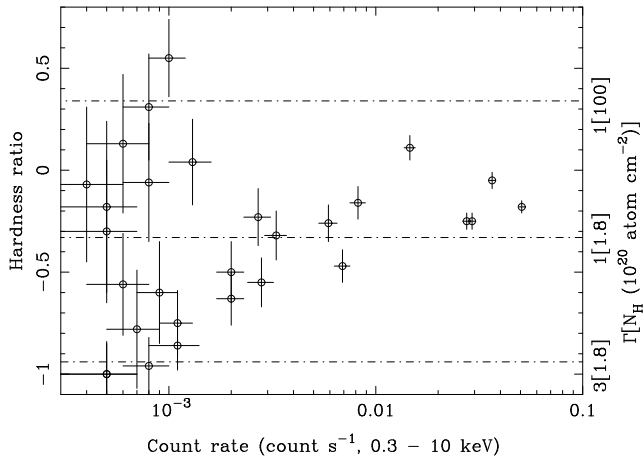}
\caption{Hardness ratio, as a function of count rate, for the sources 
detected coincident with NGC 4490.  The horizontal dot-dash lines
illustrate the dependence of HR on the spectral form, represented by
the powerlaw photon index $\Gamma$, and the intrinsic column density
(the latter in square brackets).}
\label{hrs}
\end{figure}

In order to investigate the characteristics of the detected sources
further, we calculate a hardness ratio, HR, between the soft (0.3 -- 2
keV) and hard (2 -- 10 keV) count rates for each source (see
Table~\ref{srcs}), using the hardness ratio conventions of Ciliegi et
al. (1997) where a more positive value of HR implies a harder source
spectrum.  Figure~\ref{hrs} shows the HR values of the sources
coincident with NGC 4490 plotted against source count rate.  We also
plot three fiducial markers to demonstrate the dependence of HR on
either the intrinsic spectral shape of the sources (represented here
by a powerlaw continuum with slope $\Gamma$), or the intrinsic
absorption column \nh~(note the minimum column of $1.8 \times 10^{20}
\atpcm$ corresponds to the Galactic foreground column towards NGC
4485/90 interpolated from the \hi maps of Stark et al. 1992).

Figure~\ref{hrs} demonstrates a wide scatter in HR value,
which is particularly pronounced for the faintest sources in NGC 4490.  The
observed HR values imply that the hardest sources have either negative
photon indices or very high absorption column (\nh $\gg 10^{22}
\atpcm$), whereas the softest sources must have both intrinsically low
column and a steep spectral form ($\Gamma > 3$).  We find a weighted
mean of $-0.22 \pm 0.02$ for the HR, a value which would be typical
of, for example, a luminous X-ray binary in NGC 4490 with $\Gamma =
1.6$ and an absorption column \nh $\sim 4 \times 10^{21} \atpcm$,
given that the total line-of-sight column through the disk of the
galaxy is $\sim 4 - 8 \times 10^{21} \atpcm$ (Clemens, Alexander \&
Green 1998).

\begin{table*}
\caption{The variation of HR with flux across three count rate limited 
samples of sources in NGC 4490.}
\begin{tabular}{lccccc}\hline
Sample	& Number of sources	& Count rate limits (0.3 -- 10 keV)
& Extreme HR values	& Weighted mean HR 	& Median HR \\\hline
Bright	& 5	& $> 0.013 \ctsec$	& $0.11 \rightarrow -0.25$	&
$-0.16 \pm 0.02$	& -0.18 \\
Medium	& 8	& $1.3 \times 10^{-3} \rightarrow 0.013 \ctsec$	&
$-0.16 \rightarrow -0.63$	& $-0.35 \pm 0.04$	& -0.395 \\
Faint	& 16	& $< 1.3 \times 10^{-3} \ctsec$	& $0.55 \rightarrow
-1.0$	& $-0.59 \pm 0.05$	& -0.43 \\
\hline
\end{tabular}
\label{fluxsamples}
\end{table*}

We investigate the variation of HR with flux level in NGC 4490 further
in Table~\ref{fluxsamples}.  In this, we split the sources up into
three samples, with the limits between the samples set at 0.0013 and
0.013 $\ctsec$ (0.3 -- 10 keV) respectively.  Assuming the inferred
average spectral shape for the full sample ($\Gamma = 1.6$, \nh $= 4
\times 10^{21} \atpcm$) and a distance of 7.8 Mpc, the count rate
limits convert to observed source luminosities of $< 10^{38} \ergsec$
(faint sample), $10^{38} - 10^{39} \ergsec$ (medium) and $> 10^{39}
\ergsec$ (bright).  Table~\ref{fluxsamples} shows that as the sources
get fainter, their average spectrum (indicated by both the weighted
mean\footnote{The weighted mean is calculated for each sample using
the formal statistical error on each individual hardness ratio as its
weighting factor.}  and median values) gets softer.  This suggests
that either a new, spectrally softer population of sources is
appearing at fainter fluxes, or that the fainter sources are less
absorbed than their bright counterparts.

Viable candidates for a softer source population at luminosities $\la
10^{38} \ergsec$ include X-ray bright supernova remnants (\eg Schlegel
1995) or intrinsically very soft X-ray binaries such as ``super-soft
sources'' (SSS; see below).  To examine whether a new, intrinsically
soft population is appearing at faint fluxes we have examined the four
softest sources in more detail.  They have HR values of $\la -0.9$,
which are equivalent to temperatures $< 2$ keV for a solar abundance
MEKAL thermal plasma model, consistent with the temperatures found for
supernova remnants in nearby galaxies (\cf RPS97, who find a typical
temperature of $0.36 \pm 0.2$ keV for supernova remnants in their \ro
PSPC survey of a small sample of nearby galaxies).  Unfortunately, all
four sources are too faint for spectroscopy, and even when combined we
do not have sufficient photons for detailed spectral analysis.
However, we are able to further examine their X-ray colours by
creating an 0.3 -- 0.5 keV image of the galaxy.  Three of the soft
sources do not show significant emission ($\la 1$ count) in this very
soft band, implying they are only moderately soft, or they are
absorbed by a column of $\sim 10^{21} \atpcm$.  However, all the
counts detected from the source CXOU J123039.1+413751 are seen below
0.5 keV.  This makes it a {\it bona fide\/} candidate for a super-soft
source, a class of moderately luminous ($\sim 10^{36} - 10^{38}
\ergsec$) soft X-ray emitting objects thought to contain a white dwarf
star with steady nuclear burning present in its envelope (Van den
Heuval et al. 1992).  Typically such a source has an X-ray spectrum
characterised by a black-body continuum with a peak at $\sim 50$ eV.
Many of these objects are now known in local group galaxies (Greiner
2000), and they are becoming more commonly seen as a minority
population (at the few percent level) in deep surveys of nearby
galaxies, for example in the \ro PSPC survey of M31 (Supper et
al. 2001) and in a \chan observation of the central regions of M101
(Pence et al. 2001).

We examine the second option, namely that the fainter sources are less
absorbed, by creating composite spectra of all the sources in both the
faint and medium flux samples (we examine the individual spectra of
the bright sources in the next section).  Spectra were extracted from
the cleaned event files using the {\small PSEXTRACT} script, which
also creates the appropriate instrument response files.  We used the
source region files output by {\small WAVDETECT} to define the source
areas, and then extracted all spectra simultaneously to form one
output file for each flux-limited sample.  A background file was
extracted by moving all the source regions $\sim 5$ arcseconds to the
west of their original positions, in each case into a source-free
region, and following the same extraction procedure (which is
automatically executed by {\small PSEXTRACT}).  Finally, the spectra
were grouped to 25 counts per spectral bin and analysed in {\small
XSPEC}, where we only consider spectral bins in the 0.5 - 10 keV range
in which the ACIS-S calibration is best defined\footnote{In practice
the spectra presented in this paper do not extend above 8 keV, as the
detector area falls off sharply at this energy.  Hence all fluxes
hereafter are generally quoted with 8 keV as the upper energy limit.}.

\begin{figure}
\centering
\includegraphics[width=6cm,angle=270]{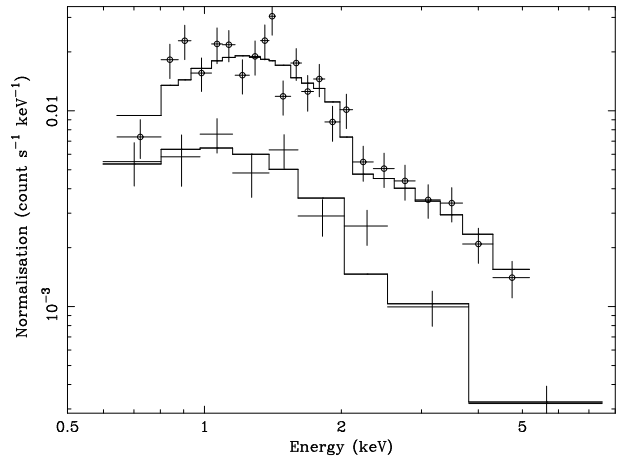}
\caption{The cumulative \chan ACIS-S spectra of the faint and medium
source samples in NGC 4490.  Both spectra are shown with their
best-fitting thermal bremsstrahlung model.  The medium sample data
points are highlighted by an open circle.}
\label{samplespec}
\end{figure}

\begin{table*}
\caption{Simple fits to the composite spectra of the flux-limited samples.}
\begin{tabular}{llcccc}\hline
Spectral model	& Sample	& \nh~($\times 10^{20} \atpcm$)	& $\Gamma/kT$ (keV)	&
Reduced $\chi^2$ 	& Degrees of freedom \\\hline
WA*PO$^a$	& Faint		& $0.19^{+0.11}_{-0.14}$	&
$1.55^{+0.31}_{-0.21}$	& 1.43	& 6 \\
	& Medium	& $0.42^{+0.12}_{-0.10}$	&
$1.78^{+0.13}_{-0.23}$	& 1.33	& 20 \\
 \\
WA*BR$^a$	& Faint		& $0.13^{+0.13}_{-0.12}$	& $> 6$	& 1.4
& 6 \\
	& Medium	& $0.36^{+0.08}_{-0.09}$	&
$5.8^{+5.9}_{-1.9}$	& 1.32	& 20 \\
 \\
WA*MEKAL$^a$$^b$ & Faint	& $0.15 \pm 0.13$	& $> 5$	& 1.4	& 6 \\
	&  Medium	& $0.33 \pm 0.07$	& $7.4^{+6.5}_{-2.1}$
& 1.43	& 20 \\
\hline
\end{tabular}
\begin{tabular}{l}
Notes:\\ $^a$The model components are: WA - cold absorption; PO -
powerlaw continuum; BR - thermal bremsstrahlung; MEKAL - \\ Mewe,
Kaastra \& Liedahl (1995) thermal plasma model.\\  $^b$ The MEKAL is assumed to
have solar abundance (see text).
\end{tabular}
\label{samplefits}
\end{table*}

The composite spectra were fit with simple models (one component plus
absorption) due to the photon-limited nature of the faint sample
spectrum ($< 250$ counts).  We show the best fitting results in
Table~\ref{samplefits}, and the data in Figure~\ref{samplespec}.  In
the table, the models are defined as per the {\small XSPEC} syntax,
and the quoted errors are the 90\% limits for one interesting
parameter (or the upper/lower limit where only one constraint could be
placed).  All models give reasonably similar fits to the data.  Note
that the MEKAL models were assumed to have solar abundance, since the
data quality is too poor (particularly in the case of the faint
sample) to distinguish thermal line emission.  It appears that a
difference in the average absorption gives rise to the change in HR
between the two samples, that is to say that the faint sample sources
are on average less absorbed than the medium sample sources.  Again,
there may be two explanations, namely that the more luminous sources
either have a higher intrinsic absorption, or are preferentially
located deeper within the galaxy disk (or in regions of denser ISM)
than the fainter sources.  The average column of \nh~$= 4
\times 10^{21} \atpcm$ for the medium sample places the absorbed
objects almost exactly half-way through the \hi cloud associated with
NGC 4490, supporting the latter of the two scenarios.  The fainter
sources, on the other hand, have a column more consistent with their
presence either on the edge of, or on the near-side of, NGC 4490.

\subsection{Ultraluminous X-ray sources}

The ``bright'' sample is of particular interest given that all its
sources have X-ray luminosities which lie above $10^{39} \ergsec$.  If
we include the source CXOU J123030.6+414142, which is coincident with
NGC 4485, this gives us six sources in the regime of the so-called
``ultraluminous X-ray sources'' (ULX)\footnote{A note on taxonomy: in
this paper we refer to the subset of non-nuclear X-ray sources in
nearby galaxies with observed X-ray luminosities in excess of $10^{39}
\ergsec$ as ultraluminous X-ray sources (ULX).  Previous papers have
variously referred to such sources as ``ultra-luminous compact X-ray
sources'' (ULX; Makishima et al. 2000), ``super-luminous X-ray
sources'' (SLS; e.g. RW2000), ``super-Eddington sources'' (SES;
e.g. Roberts et al. 2001) or ``intermediate-luminosity X-ray objects''
(IXOs; e.g. Strickland et al. 2001).} within the NGC 4485/90 system.
To place this in context, the \ro HRI survey of bright, nearby
galaxies of RW2000 detected only 28 ULX in a total of 83 nearby
galaxies (albeit in a narrower and softer energy band that is more
sensitive to absorption than that of \chan).  We can estimate the
relative overabundance of ULX in the NGC 4485/90 system, compared to
the local average, using the form of the discrete source luminosity
distribution for spiral galaxies presented in RW2000, which integrates
to $\sim 0.2$ ULX per $10^{10}$ L$_{\rm B}$ (where L$_{\rm B}$ is the
luminosity in blue light in solar units).  Taking L$_{\rm B} = 1.2
\times 10^{10}$ L$_{\odot}$ for NGC 4485/90 (Tully 1988) leads to a
prediction of $\sim 0.25$ ULX in the \ro band.  RW2000 catalogued
three ULX in NGC 4485/90 in their \ro HRI survey, implying an
overabundance factor of $\sim 12$.  Clearly, this galaxy system has an
exceptionally high incidence of ULX.  In the rest of this section we
investigate the X-ray properties of the individual ULX in the NGC
4485/90 system.  The implications of the discovery of so many ULX in
this interacting galaxy pair is discussed further in \S 7.

\subsubsection{Evidence of extension of the sources?}

Firstly, we investigate whether the ULX all appear truly point-like
under close analysis.  To define the observational appearance of a
point-like source in an on-axis \chan observation, we obtained the
images of the QSOs PG 2302$+$029 and PKS 2126-158 from the \chan
archive, which have a total of $\sim 400$ and $\sim 30000$ raw \chan
source counts respectively.  For each source we derived a radial
profile from the full band image, in single (0.492$''$) pixel steps,
out to a radius of 20 pixels, and subtracted the background using a
further annulus located between 20 and 30 pixels away from the source
position.  We fit the core of the resulting profile with a Gaussian
function, and found best fits to the FWHM of the profiles of $2.2 \pm
0.3$ and $2.03 \pm 0.03$ pixels (90\% errors) for PG 2302$+$029 and
PKS 2126-158 respectively, which are consistent with the values of
$\sim 1.5 - 2$ pixels measured in the on-ground calibration phase (see
the calibration pages on {\tt http://cxc.harvard.edu}).  We repeated
the procedure for each of the six brightest sources in the NGC 4485/90
system.  All of the sources were found to have radial profiles
consistent (within the 90\% errors) with those found for the QSOs.  We
therefore conclude that none of the bright sources appears to possess
an extended core.

As a further test of extent, we examined the possibility of low-level
enhancements in the wings of the PSF produced by a faint, extended
component.  To do this we compared the number of counts within the
annuli between 6 and 20 pixels away from each source to the total
counts within the 20 pixel radius (note that the 90\% encircled energy
radius for \chan is $\sim 4$ pixels).  In the case of the QSOs we
found that $\sim 1 - 3\%$ of the total counts were found within this
region.  Once nearby sources were excluded, this figure was again
replicated for each of the bright source sample, with one exception.
Source CXOU J123036.3+413837 has $\sim 4\%$ of the counts within 20
pixels in the 6 -- 20 pixel annulus.  Inspection of the data shows
that this is not necessarily due to an extension to the PSF for the
source, but that it is more likely to be due to the location of the
source in enhanced diffuse emission close to the galaxy nucleus.  We
discuss this diffuse emission further in \S 5.  We therefore conclude
that there is no evidence for extension in any of the bright sources,
and so all are considered to be individual discrete sources at the
resolution of \chan ($\sim 0.5''$, which is equivalent to $\sim 20$ pc
at the distance of NGC 4485/90).

\subsubsection{X-ray spectra}

\begin{figure*}
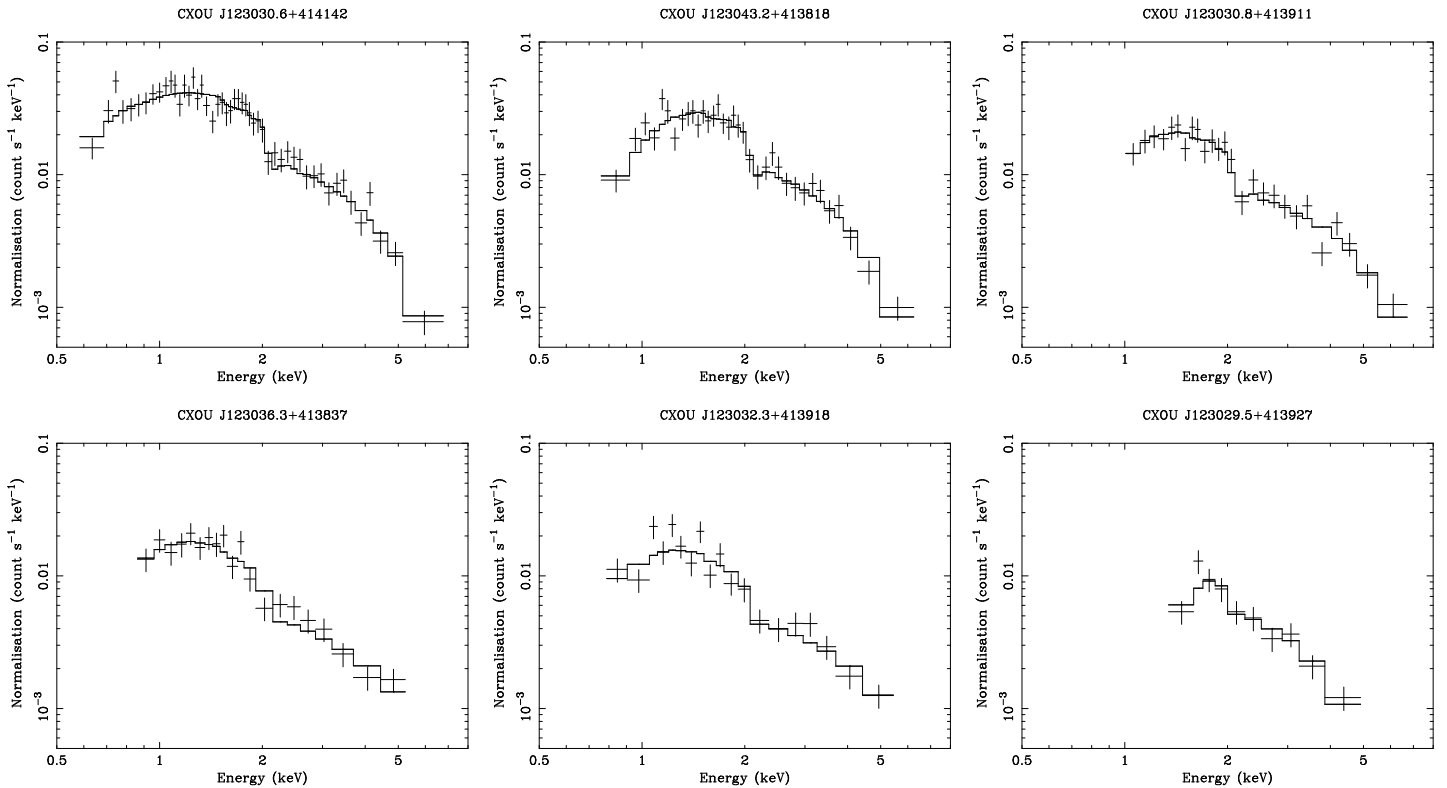

\centering
\includegraphics[width=5cm,angle=270]{fig4a.ps}\hspace*{0.2cm}
\includegraphics[width=5cm,angle=270]{fig4b.ps}\hspace*{0.2cm}
\includegraphics[width=5cm,angle=270]{fig4c.ps}\vspace*{0.3cm}
\includegraphics[width=5cm,angle=270]{fig4d.ps}\hspace*{0.2cm}
\includegraphics[width=5cm,angle=270]{fig4e.ps}\hspace*{0.2cm}
\includegraphics[width=5cm,angle=270]{fig4f.ps}
\caption{\chan ACIS-S spectra (with best-fitting models) of the six
ULX in the NGC 4485/90 system.}
\label{spectra}
\end{figure*}

The X-ray spectrum of each source was extracted using the {\small
PSEXTRACT} script, in each case using a 6-pixel radius aperture about
the source position and a local background region.  The spectra were
again fit with simple (absorption plus a single component) models in
{\small XSPEC}, namely absorbed powerlaw continua, thermal
bremsstrahlung and multi-colour disk black-body (hereafter MCD BB)
models.  The last of these models is found to be the best fit to many
ULX by Makishima et al. (2000), and describes the expected spectrum of
an optically-thick accretion disk in a black-hole binary system in a
soft (high) state as calculated by Mitsuda et al. (1984).  These
models gave statistically good fits to the data in almost all cases.
We also investigated absorbed black-body continua and optically-thin
thermal components (MEKAL model), but none produced adequate fits to
the data, with one exception (see below).  We detail the fits in
Table~\ref{xspecfits} in descending order of the observed ULX count
rate, in each case highlighting a ``best-fit'' model used primarily to
calculate the luminosity of each source.  We detail the motivation for
the choice of the ``best-fit'' model in the following paragraphs.  The
data and best-fitting models for all of the sources are shown in
Figure~\ref{spectra}, plotted on identical scales for direct
comparison.

\begin{table*}
\caption{X-ray spectral fits to the six ULX in NGC 4485/90.}
\begin{tabular}{lcccccccccc}\hline
Source	& \multicolumn{3}{c}{WA*PO$^a$}	& \multicolumn{3}{c}{WA*BR$^a$}	&
\multicolumn{3}{c}{WA*DISKBB$^a$}	& \lx$^b$ \\
(CXOU J)	& \nh$^c$	& $\Gamma$	& $\chi^2$/dof	&
\nh$^c$	& $kT^d$ & $\chi^2$/dof	& \nh$^c$	& $kT_{\rm in} ^e$
& $\chi^2$/dof & \\\hline 
123030.6$+$414142	& $0.37 \pm 0.06$	&
$1.83^{+0.14}_{-0.12}$	& 65/47	& $0.31^{+0.09}_{-0.05}$	&
$5.4^{+2.0}_{-1.0}$	& 56/47	& $0.19 \pm 0.04$	&
$1.40^{+0.14}_{-0.12}$	& {\bf 46/47}	& 4.0 (4.6) \\
 \\
123043.2$+$413818	& $0.94^{+0.17}_{-0.14}$	&
$2.22^{+0.22}_{-0.20}$	& 40/32	& $0.78^{+0.12}_{-0.11}$	&
$3.5^{+1.1}_{-0.7}$	& 33/32	& $0.59^{+0.11}_{-0.09}$	&
$1.19^{+0.14}_{-0.12}$	& {\bf 28/32}	& 3.1 (4.4) \\
 \\
123030.8$+$413911	& $0.78^{+0.28}_{-0.24}$	&
$1.80^{+0.31}_{-0.28}$	& {\bf 22/23}	& $0.62^{+0.21}_{-0.19}$	&
$7.9^{+10.7}_{-2.3}$	& 23/23	& $0.38^{+0.19}_{-0.17}$	&
$1.71^{+0.52}_{-0.30}$	& 26/23	& 2.9 (4.6) \\
 \\
123036.3$+$413837	& $0.43 \pm 0.18$	&
$1.80^{+0.30}_{-0.31}$	& {\bf 19/17}	& $0.33^{+0.13}_{-0.14}$	&
$6.6^{+14.4}_{-1.8}$	& 19/17	& $0.17^{+0.14}_{-0.12}$	&
$1.47^{+0.44}_{-0.28}$	& 20/17	& 1.9 (2.6) \\
 \\
123032.3$+$413918	& $0.47^{+0.19}_{-0.14}$	&
$1.73^{+0.30}_{-0.28}$	& {\bf 24/16}	& $0.39^{+0.14}_{-0.11}$	&
$7.6^{+18.0}_{-1.9}$	& 24/16	& $0.24^{+0.13}_{-0.11}$	&
$1.59^{+0.40}_{-0.33}$	& 24/16	& 1.9 (2.6) \\
 \\
123029.5$+$413927	& $3.6^{+1.4}_{-1.1}$	&
$3.8^{+1.0}_{-0.9}$	& 5/7	& $2.7^{+0.9}_{-0.8}$	&
$1.5^{+1.1}_{-0.4}$	& {\bf 6/7}	& $2.1^{+0.9}_{-0.7}$	&
$0.8^{+0.3}_{-0.1}$	& 7/7	& 1.0 (4.9) \\\hline
\end{tabular}
\begin{tabular}{l}
Notes:\\ $^a$ spectral model components as per Table~\ref{samplefits}, except
DISKBB which is the {\small XSPEC} tabulation of the multi-colour disk 
black-body model.\\
$^b$ observed luminosity in the 0.5 - 8 keV band in units of
$10^{39} \ergsec$.  Figures in brackets give the intrinsic
(unabsorbed) luminosity.\\  We calculate these values using the
best-fit model
highlighted by showing its $\chi^2$/dof in bold face.\\ $^c$
absorption column in units of $10^{22} \atpcm$.\\ $^d$ temperature in
keV.\\ $^e$ inner accretion disk temperature in keV
\end{tabular}
\label{xspecfits}
\end{table*}

All the sources display absorption well in excess of the foreground
Galactic \los, and values are consistent with the sources being
embedded within the \hi cloud that envelopes the NGC 4485/90 system.
The five brightest sources are all well-fit by the MCD BB with a
temperature consistent with that found for other ULX by Makishima et
al. (2000) (i.e.  $kT_{\rm in}$ in the range 1.1 -- 1.8 keV).  Indeed,
this model is clearly the best fit for the two brightest sources,
implying that they are plausible ultraluminous black hole X-ray binary
systems in the high (soft) state.  The next three brightest are
equally well-fit by powerlaw continua models, so whilst they are also
very likely to be ultraluminous black hole X-ray binary systems, their
actual state (high or low) is somewhat ambiguous on the basis of this
data.  This interpretation is only questionable in one source, CXOU
J123032.3$+$413918, which has an X-ray spectrum which is also well-fit
by a solar-abundance MEKAL thermal plasma model ($kT =
7.0^{+7.4}_{-2.2}$ keV).

The spectrum of the final source, CXOU J123029.5$+$413927, is the
least well-defined of the sample as it is by far the faintest of the
six sources.  Despite this, it does appear quite different to the
other five sources, with an extremely high absorption column of $> 2
\times 10^{22} \atpcm$ and also a soft underlying continuum form ($kT
\sim 1.5$ keV in the thermal bremsstrahlung model).  Its fit to the
MCD BB model actually shows a temperature below the normal range for
ULX.  A key piece of evidence as to its nature is the exact
coincidence of this source with a radio source, FIRST
J123029.4+413927.  These properties are consistent with an X-ray and
radio luminous supernova remnant (providing the motivation for the
choice of a thermal bremsstrahlung model as ``best-fit'', as opposed
to the non-thermal powerlaw continuum, in Table~\ref{xspecfits}).
Interestingly, if it is the result of a very recent supernova, the
high obscuring column could be the reason that it was not seen in the
optical regime.  We also note that the high X-ray luminosity and
absorption column are consistent with the scenario of a supernova
exploding into a dense circumstellar environment and emitting copious
X-rays due to the interaction of the supernova shock wave with the
circumstellar matter (\eg Fabian \& Terlevich 1996; Chevalier \&
Fransson 1994).

\subsubsection{Short-term X-ray variability}

We investigated the variability of each of the six ULX over the 19.5
ks observation by creating lightcurves binned to an average of 25
counts per bin, i.e. a signal-to-noise ratio of 5.  We then tested the
variability of each source against the hypothesis that the source flux
was constant using a $\chi^2$ test.  We found that none of the sources
showed any evidence for short-term variability, with each lightcurve
well-fit by a constant flux.  A visual inspection of the lightcurves
verified this result.

\subsubsection{Long-term X-ray variability}

\begin{figure*}
\centering
\includegraphics[width=12cm,angle=270]{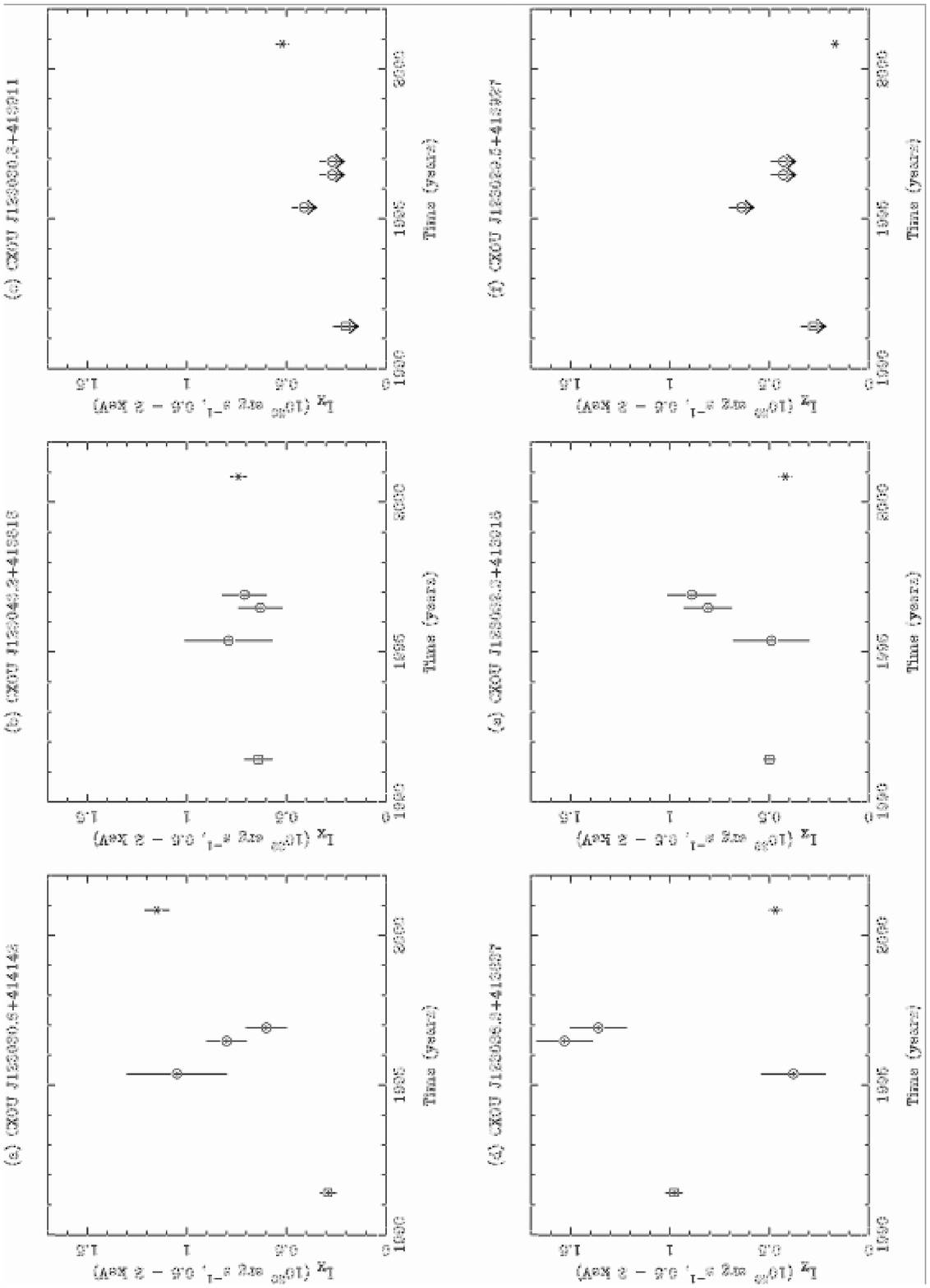}
\caption{Long-term lightcurves of the six ULX from \ro and
\chan data, in the 0.5 - 2 keV band.  We mark the \ro PSPC data point
by an open square, the three HRI points by open circles and the \chan
ACIS-S data point by an asterisk.  Upper limits are shown by the
downwards arrows.}
\label{ltlcs}
\end{figure*}

The NGC 4485/90 system was observed by \ro four times during the
1990s, once with the PSPC and three times with the HRI.  We are
therefore able to extract information on the long-term variability of
the ULX, over a baseline of ten years, from the combined datasets.  We
use \ro PSPC count rates from the RPS97 analysis of a 30 ks
observation of the system.  They detect four of the six bright sources
(CXOU J123030.8$+$413911 and CXOU J123029.5$+$413927 are not
detected); however the moderate intrinsic spatial resolution of the
\ro PSPC, which is in the range $15 - 45''$, means that the count
rates of the sources are open to contamination by nearby sources
and/or extended emission components.  We attempt to quantify this
effect by examining the soft band \chan image.  The additional
emission above the nominal off-galaxy background within 45$''$ of the
source positions is negligible ($\ll 10\%$ of the source counts) for
both CXOU J123030.6+414142 and CXOU J123043.2+413818.  However, we
find that only 33\% of the counts from within 45$''$ of CXOU
J123032.3+413918 and 45\% of the counts within 45$''$ of CXOU
J123036.3+413837 come from each source respectively.  Our best
estimate of the count rate of each of these sources at the epoch of
the \ro PSPC observation is therefore the observed count rate
multiplied by this fraction.  The same consideration is not necessary
in the case of the \ro HRI datasets, where the bright sources are
fully resolved from their neighbouring components by the $\sim 5''$
beam.  We take the count rates for the latter two HRI observations
from RW2000, and perform a new analysis to derive the count rates from
the first \ro HRI observation (observation ID rh600697n00).

We derive count rate to flux conversions for both the PSPC and HRI
detections in {\small XSPEC}, using the best fit models found in the
spectral analysis of the \chan data.  These are folded through the
relevant instrument response matrices, and then normalised to the flux
at each epoch using the predicted and observed count rates for each
instrument.  Finally, we convert the fluxes into luminosities for the
NGC 4485/90 system.  The upper limits on the luminosities of the two
sources not detected by \ro were calculated in a similar fashion,
except that we set a conservative upper limit on the count rate at the
flux of the faintest (normally $\sim 3\sigma$ significance) source
detection count rate in each field.  The results of these calculations
are shown in Figure~\ref{ltlcs}, where we present a long-term
lightcurve for each of the six ULX.

As can be seen, four sources are detected in each observation, and two
sources are only detected in the \chan ACIS-S observation.  The
reasons for the non-detections differ; CXOU J123030.8$+$413911 appears
to be transient in nature, with an observed (0.5 -- 2 keV) luminosity
in the \chan data epoch well in excess of the limits placed on it from
the earlier measurements, whereas CXOU J123029.5$+$413927 is faint in
the 0.5 -- 2 keV band due to its highly absorbed spectrum and so,
assuming a constant flux, is undetectable in the previous observations
(though, of course, this data cannot rule out variability in this
source).  Of the four detections, three sources appear strongly
variable over the 10 year baseline (factors $> 2$), with their
intrinsic flux increasing and declining throughout this period,
characteristic of accreting sources.  One source, however, (CXOU
J123043.2+413818) does not behave in this way.  Instead, it shows a
reasonably constant luminosity of $\sim 7 \times 10^{38} \ergsec$ (0.5
- 2 keV) over the 10-year baseline, implying that if it is an X-ray
binary then it might have a duty cycle with a long, remarkably stable
epoch in the high state.

\section{Diffuse emission}

\begin{figure*}
\centering
\includegraphics[width=12cm,angle=270]{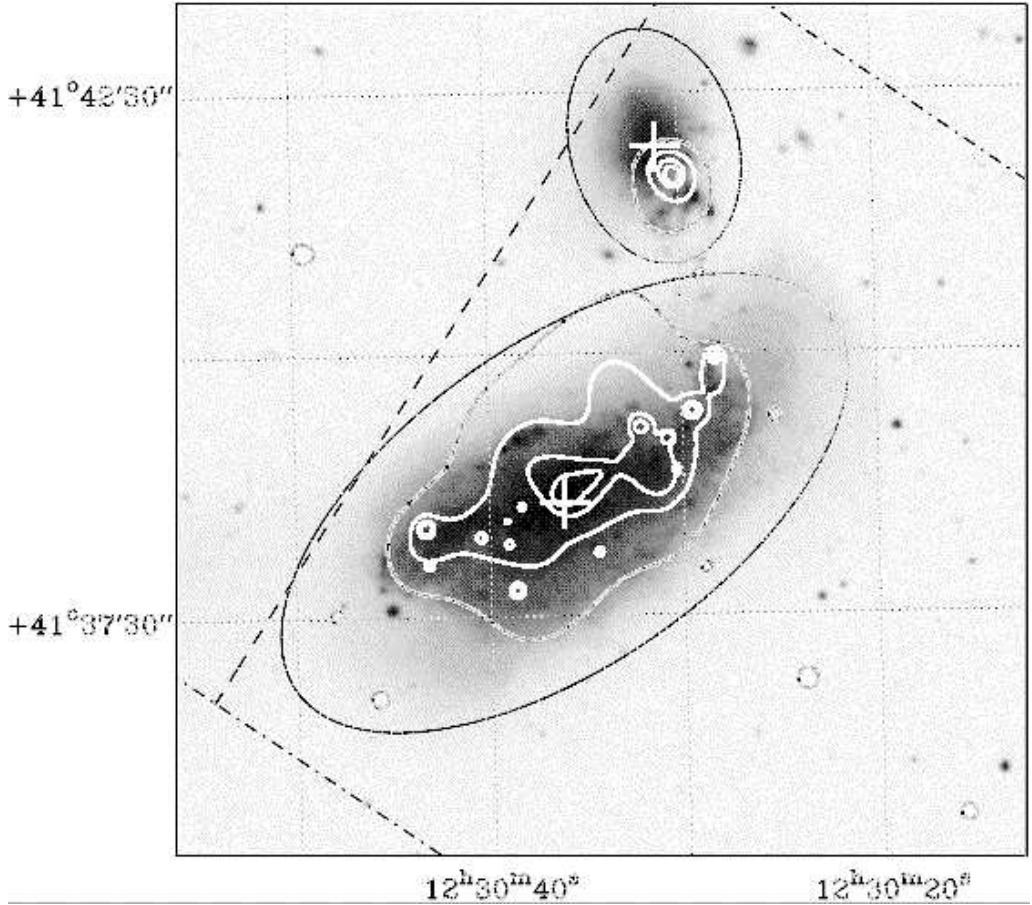}
\caption{Contours from the {\small CSMOOTH}ed soft band image of NGC
4485/90 overlaid onto the DSS-2 blue image.  Contours are marked at
2.5 (black), 5, 10 and 20 times the nominal off-galaxy surface
brightness.  \d25 ellipses, nuclear positions and the edges of the
ACIS-S chips are shown as per Figure~\ref{dss2srcs}.}
\label{diffuse}
\end{figure*}

An inspection of the raw \chan data suggests that there may be a
diffuse emission component present in and around the centre of NGC
4490.  We investigated this by adaptively smoothing the \chan images
using the {\small CSMOOTH} routine.  This provides the user with a
image that has been smoothed using a 2-D Gaussian kernel, with a
variable size that is set to increase until it provides a certain
statistical acceptability (in this case at least 3$\sigma$) for the
count rate at each point in the image.  These images confirm the
presence of widespread diffuse emission within NGC 4490 and, to a
lesser extent, NGC 4485.  We show a contour map of the {\small
CSMOOTH}ed soft (0.3 -- 2 keV) band data overlaid on the DSS2 blue
image of the galaxy pair in Figure~\ref{diffuse}.  The positions of
the brightest point sources as well as the diffuse component are
shown, but the variable kernel size and the tight \chan PSF mean that
the point source flux is not significantly smoothed into the
surrounding emission.  Hence away from the source positions we are
seeing a truly separate component.

Figure~\ref{diffuse} shows that the diffuse emission associated with
NGC 4490 has two main peaks; the brightest is centred immediately to
the north-west of the nuclear region, and a secondary peak lies $\sim
1$ arcminute to the west, just to the south of the ULX CXOU
J123030.8$+$413911 and CXOU J123032.3+413918.  The position of the
nuclear peak matches the peak in both 100$\mu$m emission and \Ha~
noted by Thronson et al. (1989), and the secondary peak matches an
extension to the west of the nucleus in the same emission bands.  The
overall distribution of the diffuse emission roughly follows the
optical light distribution on the southern side of NGC 4490.  However,
the diffuse emission contours (and in particular the lowest contour)
are extended away from the high optical surface brightness regions to
the north.  This may imply that the diffuse component in this region
of the galaxy is perturbed by high energy processes, e.g. outflows
and/or superwinds originating in the star formation regions.  There is
an additional enhancement in the diffuse X-ray surface brightness in
the southern half of NGC 4485, seemingly centred around the bright
X-ray source CXOU J123030.6+414142 in the tidal tail of the galaxy.
However, the asymmetric shape and size of the emission region implies
that it is not simply the wings of the source PSF, but that it is a
true detection of a region of diffuse emission.  The presence of a
possible hot ISM in just the southern tidal tail of NGC 4485 agrees
with the observations of Clemens, Alexander \& Green (2000), where the
atomic, molecular and dusty gas associated with the galaxy are largely
located in the same region, and a bow-shock is identified in the \hi
gas to the north of this region.  This is consistent with a scenario
in which the motion of NGC 4485 in a northerly direction through the
extended \hi cloud of NGC 4490, at an estimated velocity of 90 km
s$^{-1}$, causes NGC 4485 to be stripped of all components of its ISM
through the process of ram-pressure stripping (Clemens, Alexander \&
Green 2000).

We performed the same analysis for the hard (2 -- 10 keV) band, but
this band was devoid of significant amounts of unresolved emission,
with none appearing in the {\small CSMOOTH}ed image above a level of
twice the nominal off-galaxy surface brightness.

As a next step, we investigated the X-ray spectrum of the diffuse
emission.  Firstly, we excluded all photons within the discrete source
data extraction regions (as used in previous sections) from the event
list.  We then extracted the spectrum of the remaining emission in an
elliptical region of dimensions $138''$ (semi-major axis) by $62''$
(semi-minor axis), centred on the western tip of the highest emission
contour in the centre of NGC 4490 in Figure~\ref{diffuse}, and at a
position angle of 120$^{\circ}$ east of north.  This covers most of
the area within the 2.5 times background level contour, and all the
area under the 5 times background level contour, in
Figure~\ref{diffuse}.  We note that a mere $\sim 1\%$ of the total
area covered by the ellipse is excluded by the subtraction of the
point sources, hence no correction is required for the area lost.
Finally, a background spectrum was extracted from a source-excluded
region to the south-west of the galaxy of equivalent size to the
ellipse.  The resulting spectrum of the diffuse emission, containing a
total of $\sim 1600$ counts, was again analysed in {\small XSPEC}.

\begin{figure}
\centering
\includegraphics[width=6cm,angle=270]{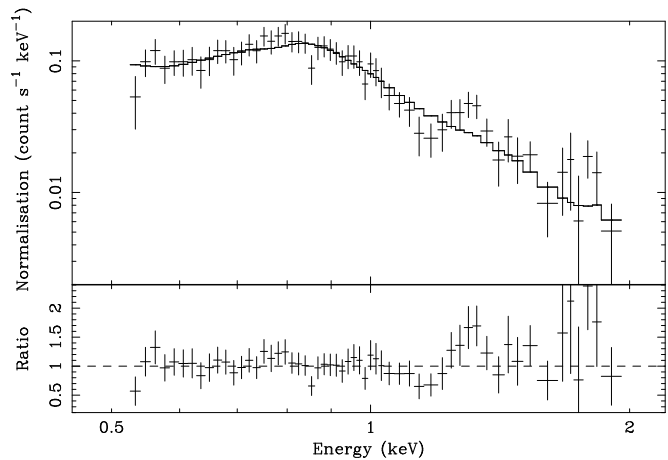}
\caption{The \chan ACIS-S spectrum of the diffuse emission in NGC
4490.  The top panel shows the data and the best fit MEKAL model,
whilst the bottom panel shows the ratio of the data divided by the
model.}
\label{despec}
\end{figure}

Due to the soft nature of the emission, and the rising ACIS-S
instrument background as a function of energy, noise dominated the
spectral data at energies above 2 keV and hence the analysis was
restricted to the 0.5 -- 2 keV range.  A variety of absorbed single
component spectral models were fit to the data, as per \S 4.2.2.  The
data were clearly best fit by an absorbed MEKAL optically-thin thermal
plasma model, with a $\chi^2$ of 43 for 53 degrees of freedom.  This
compares to $\chi^2$ of 74 - 98 (for the same degrees of freedom)
using absorbed powerlaw continuum, thermal bremsstrahlung and MCD BB
models.  The best fitting parameters are shown in Table~\ref{defit},
and the spectral data with the best fitting model is displayed in
Figure~\ref{despec}.

Line-like residuals are clearly still present in the spectrum,
particularly at around 1.3 keV.  To investigate whether these
residuals were better fit by more complicated models we fit the data
with combinations of two physical models plus absorption.  Both an
absorbed MEKAL plus highly absorbed powerlaw (WA*[MEKAL+WA*PO] in
{\small XSPEC} syntax) and an absorbed two-temperature MEKAL
(WA*[MEKAL+WA*MEKAL]) model achieved similar statistical acceptability
to the single component MEKAL fit ($\chi^2$ of $\sim 43$ for 50
degrees of freedom).  However, the presence of additional parameters
in the fit meant that we were unable to tightly constrain the values
of the fit parameters, hence we only quote the values of the
well-constrained single component fit in Table~\ref{defit}.  Both
models were highly flux-dominated by the softer MEKAL component (with
$kT \sim 0.56$ keV, abundance $\sim 0.05$ solar, absorption $5 - 10
\times 10^{20} \atpcm$) with the harder component only contributing
significantly above 1 keV (MEKAL with $kT \sim 0.84$ keV, same
abundance, extra $5 \times 10^{20} \atpcm$ absorption; or a $\Gamma
\sim 1.5$ powerlaw continuum absorbed by a column of $\sim 1.5 \times
10^{22} \atpcm$).  Hence even in the more complicated models the
spectrum is dominated by a soft thermal plasma component.  Indeed, in
the MEKAL plus powerlaw model 95\% of the flux is supplied by the
MEKAL model, emphasizing that there appears little contribution to the
diffuse emission originating in non-thermal sources (i.e. unresolved,
point-like sources).  The spectral models are consistent with the lack
of a strong diffuse component in the hard band images, and together
they suggest that the diffuse emission component seen in NGC 4490 is
predominantly thermal in nature.

\begin{table}
\caption{Best fitting model to the diffuse emission.}
\begin{tabular}{lcc}\hline
Model component	& Parameter	& Value \\\hline
WABS$^a$	& \nh ($\times 10^{20} \atpcm$)	& $5.1^{+6.0}_{-3.3}$ \\
MEKAL	& $kT$ (keV)	& $0.64^{+0.05}_{-0.1}$ \\
	& Abundance (solar units)	& $0.05 \pm 0.02$ \\\hline
\end{tabular}
\begin{tabular}{l}
Notes:\\
$^a$ Minimum column constrained to the foreground Galactic of \\$1.8
\times 10^{20} \atpcm$
\end{tabular}
\label{defit}
\end{table}

All the inferred spectral fits have absorption columns in excess of
$\sim 5 \times 10^{20} \atpcm$, over twice the Galactic \los column,
implying that the unresolved X-ray emission is coming from a component
that is at least in part mixed in with the cooler, absorbing gas in
NGC 4490.  The derived temperature of the diffuse emission is fairly
typical of the diffuse component seen in other moderate-activity
galaxies (Ptak et al. 1999), as is the significantly sub-solar
abundance.  Low ($< 10\%$ solar) abundances are generally attributable
to the thermal plasma being in a complicated, multi-temperature state
poorly described by a single temperature model (see e.g. Dahlem,
Weaver \& Heckman 1998).  The presence of line-like residuals, even
after fitting the MEKAL model, may also imply that the relative
abundances of the heavy elements are not at the solar ratio.  The best
fit spectrum has an observed 0.5 -- 2 keV flux of $2.1 \times 10^{-13}
\ergcms$, which implies a diffuse emission luminosity of $1.5 \times
10^{39} \ergsec$.

\section{Luminosity issues}

\subsection{The total X-ray luminosity of the system}

In Table~\ref{totallx} we present a decomposition of the X-ray
luminosity of the galaxy into its resolved (discrete source and
diffuse emission) parts.  The discrete source luminosities are derived
from the best fit models described in \S 4, in the 0.5 -- 8 keV range.
Intrinsic luminosities refer to the luminosity derived for each
spectral model in that band if the line-of-sight absorption is
removed.  The unresolved emission from NGC 4490 was also taken from
its best fit spectrum (c.f. \S 5), and is extrapolated here into the
wider 0.5 -- 8 keV band (though note that this increases the
luminosity by $< 10\%$ due to its intrinsically soft nature).  We
estimated the luminosity of the unresolved component in the southern
part of NGC 4485 in the following manner.  We extracted an image in
the 0.5 -- 8 keV band in a circular aperture, 30$''$ in radius, about
a position $\sim 7''$ south of CXOU J123030.6+414142, and excluding a
radius of 6 pixels around the source.  A similar-sized background
region was extracted in a source-free region $\sim 1'$ to the
south-west.  The difference between the number of counts in each image
gave an estimate of the number of counts in the diffuse component,
namely $160 \pm 18$ counts.  A similar analysis shows that $\sim 1600$
counts lie in the comparable unresolved emission component for NGC
4490.  Assuming a similar spectral form for the diffuse emission, we
derive the luminosities for the NGC 4485 diffuse component shown in
Table~\ref{totallx}.

\begin{table}
\caption{The contribution of the resolved components to the total X-ray
luminosity of NGC 4485/90}
\begin{tabular}{lcc}\hline
Component	& Observed L$_{\rm X}$	& Intrinsic L$_{\rm X}$ \\
	& \multicolumn{2}{c}{($\times 10^{39} \ergsec$, 0.5 - 8 keV)}
\\\hline
\multicolumn{3}{l}{\bf Discrete X-ray sources} \\
CXOU J123030.6+414142	& 4.0	& 4.6 \\
CXOU J123043.2+413818	& 3.1	& 4.4 \\
CXOU J123030.8+413911	& 2.9	& 4.6 \\
CXOU J123036.3+413837	& 1.9	& 2.6 \\
CXOU J123032.3+413918	& 1.9	& 2.6 \\
CXOU J123029.5+413927	& 1.0	& 4.9 \\
Medium sample 		& 2.3	& 3.2 \\
Faint sample 		& 0.7	& 0.8 \\
 \\
\multicolumn{3}{l}{\bf Diffuse emission} \\
NGC 4485		& 0.2	& 0.4 \\
NGC 4490		& 1.6	& 3.6 \\
 \\
{\bf Total} 		& {\bf 19.6}  & {\bf 31.7}  \\\hline
\end{tabular}
\label{totallx}
\end{table}

The total observed X-ray luminosity of the galaxies, of $\sim 2 \times
10^{40} \ergsec$, is clearly dominated by the discrete X-ray sources.
In fact, the diffuse emission contributes no more than $\sim 10\%$ to
the observed X-ray luminosity in the 0.5 -- 8 keV range.  The discrete
source population is dominated by the very brightest sources, with the
brightest three sources contributing over half of the total observed
flux.

The only previous X-ray study of NGC 4485/90 to separate out the
discrete X-ray sources from a diffuse component was the \ro PSPC study
of RPS97.  They found a total 0.1 -- 2 keV luminosity of $1.2 \times
10^{40} \ergsec$ for NGC 4490, with $(55 \pm 5)\%$ of the total flux
originating in a diffuse component.  This diffuse luminosity is well
in excess of what we observe for the system, which is obviously in
part due to the large number of additional sources we resolve in the
system.  However, a large part of the discrepancy may come from the
more extensive diffuse emission component detected by RPS97 (see their
Figure 13).  It is possible that we do not detect this component due
to a combination of its low surface brightness and a very soft
intrinsic spectral form, the latter implying that it is mainly evident
at energies below the nominal {\it CHANDRA\/} detector response.  This
``very soft'' emission may represent a hot gas halo surrounding the
galaxy pair.

\subsection{The luminosity function of the discrete X-ray sources}

\begin{figure}
\centering
\includegraphics[width=6cm,angle=270]{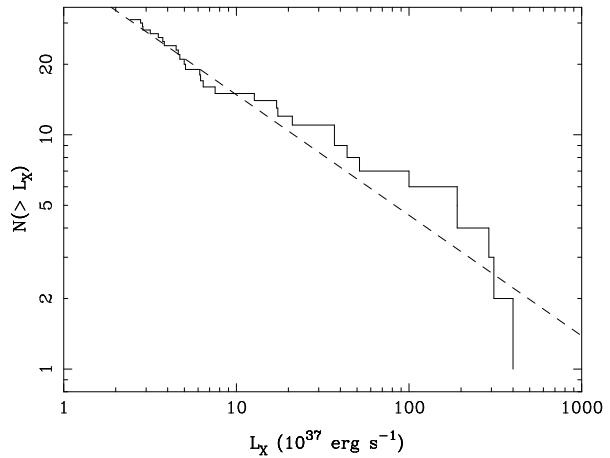}
\caption{The luminosity function of the discrete X-ray sources in the
NGC 4485/90 system.  The cumulative luminosity function is shown as
the solid line, and the best fit (powerlaw) relationship is shown as a
dashed line.  Luminosities are calculated as per the text.}
\label{lumdist}
\end{figure}

With the advent of \chan we are able to perform highly sensitive
studies of the point source populations in a wide variety of nearby
galaxies for the first time.  Luminosity functions are an increasingly
popular diagnostic of these X-ray source populations, as they provide
an empirical, and readily comparable, view of the numbers and relative
luminosities of sources within such systems.  The luminosity function
for the discrete X-ray source population of the NGC 4485/90 system is
shown in Figure~\ref{lumdist}.  We calculate this luminosity function
by extrapolating our fluxes into the 0.3 -- 8 keV band, allowing
direct comparison with the various luminosity functions presented by
Kilgard et al. (2002).  The luminosities of the ULX were taken from
their best fit spectral models; sources with count rates in the
``medium'' and ``faint'' flux regimes had their 0.3 -- 10 keV count
rates converted to 0.3 -- 8 keV fluxes using the best fit model to the
composite source spectrum in each flux regime (c.f. \S 4.1).  The
derived source luminosities cover a range of two orders of magnitude,
from $\sim 2 \times 10^{37} - 4 \times 10^{39} \ergsec$.

The slope of the luminosity function was determined to be $-0.57 \pm
0.1$ using the maximum likelihood statistic of Crawford, Jauncey \&
Murdoch (1970).  We note that this slope is not heavily dependent upon
the spectral models used in calculating the count-rate to flux
conversion: for instance, if we use a 5 keV thermal bremsstrahlung
model with a Galactic line-of-sight column, we derive a
fully-consistent slope of $-0.59 \pm 0.1$.  There is a hint of a
steepening in the luminosity function slope at the high luminosity
end, at $\sim 2 \times 10^{39} \ergsec$, though this cannot be
interpreted as a significant break feature since only four data points
lie beyond the putative break luminosity.  We note that breaks in the
luminosity functions of several other galaxies have been observed in
\chan observations (e.g. NGC 4697, Sarazin et al. 2000; NGC 1553,
Blanton et al. 2001), where breaks at a few $\times 10^{38} \ergsec$
have been attributed to the transition between neutron star and black
hole X-ray binaries.  Another possibility is that the break energy is
variable, and dependent upon the star formation history of the galaxy
(Wu 2001; Kilgard et al. 2002).  In the case of NGC 4485/90 the high
luminosity of the possible break and the active star formation would
favour the latter explanation, especially when it is considered that
the assumed distance of the galaxies would have to be overestimated by
a factor at least two to be consistent with a the putative break being
at a few $\times 10^{38} \ergsec$.

Kilgard et al. (2002) detail the difference in luminosity function
slopes between different types of galaxies.  In particular, there
appears to be an evolution in the steepness of the luminosity function
between starburst systems (which have relatively flat slopes, $\sim
-0.5$), normal spiral galaxies (slope $\sim -1.1$) and early-type
galaxies (slope $\sim -1.7$).  Kilgard et al. show this slope is
correlated with the 60$\mu$m luminosity of each galaxy, which is a
measure of the star formation activity in each galaxy.  Hence, the
luminosity function slope is related to the age of the stellar
population in each system, with the young stellar populations in
starburst galaxies showing a flatter slope as a direct consequence of
hosting a relatively larger number of high luminosity X-ray binary
systems (probably high-mass X-ray binaries) than older stellar
populations.  A similar conclusion is drawn by Tennant et al. (2001)
in comparing the luminosity functions of the disk and bulge of M81.
The luminosity function slope we derive for NGC 4485/90 places it
firmly in the starburst regime; indeed, a value of $-0.57 \pm 0.1$ is
a close match to that of the Antennae.  The relatively shallow
luminosity function slope therefore implies that the NGC 4485/90
system hosts a young X-ray source population, similar to that observed
in classical starburst systems.

An useful application of the derived luminosity function is to
interpolate it below the minimum detected luminosity, in order to
place limits on the contribution of an unresolved, faint source
population to the the observed diffuse luminosity of the galaxies.
Assuming the luminosity function slope remains unchanged below $2
\times 10^{37} \ergsec$, we derive a total luminosity of $\sim 1.2
\times 10^{39} \ergsec$ for an unresolved X-ray source population.
This compares to an observed diffuse X-ray luminosity of $1.8 \times
10^{39} \ergsec$, implying the unresolved source population may
contribute well over half the diffuse X-ray flux.  However, we have
shown in previous sections that this diffuse component appears
predominantly thermal in nature.  This apparent dichotomy may be
explained if either the source population below $2 \times 10^{37}
\ergsec$ is dominated by discrete thermal X-ray sources
(i.e. supernova remnants), or if a further break is present in the
luminosity function slope at or close to $10^{37} \ergsec$ which
results in a flattening of the slope at lower source luminosities.

\subsection{NGC 4485/90 as an X-ray starburst system}

There are many indications that NGC 4485/90 should be classified as a
starburst system.  For instance, RPS97 classify NGC 4490 as a
starburst on the basis of its high L$_{FIR}$/L$_B$ ratio ($> 0.38$)
and its high FIR colour temperature ($S_{60\mu m}/S_{100\mu m} =
0.52$).  Its current star formation rate is also similar to that of
the prototypical galaxy merger starburst in the Antennae, which is
$\ga 5$ M$_{\odot}$ yr$^{-1}$ (Stanford et al. 1990) (c.f. 5
M$_{\odot}$ yr$^{-1}$ in NGC 4490; Clemens, Alexander \& Green 1999).
It does not, however, show the classical X-ray characteristics of a
compact starburst system such as M82 or NGC 253, with in particular no
evidence of a Galactic-scale superwind emanating from the nucleus of
NGC 4490 in X-ray observations (at best we see a diffuse X-ray halo
that may be extended to the north of the galaxy; see \S 5).

\begin{table}
\centering
\caption{A comparison of the X-ray luminosity diagnostics in NGC
4485/90, the Antennae and NGC 3256.}
\begin{tabular}{lccc}\hline
Galaxy	& L$_{FIR}$/L$_B$$^a$	& L$_{\rm X, diff}$$^b$	& L$_{\rm X,
src}$$^c$ \\\hline
NGC 4485/90	& 0.6	& 0.2	& 1.8 \\
Antennae	& 0.8	& 3.8	& 4.6 \\
NGC 3256	& 10	& $\sim 28$	& $\sim 7$ \\\hline
\end{tabular}
\begin{tabular}{l}
Notes:\\
$^a$ Based on FIR fluxes from Soifer et al. (1989), and blue\\
magnitudes from de Vaucouleurs et al. (1991).\\  $^b$
Diffuse X-ray luminosity in units of $10^{40} \ergsec$,\\ 0.5 - 8 keV.\\
$^c$ Combined source X-ray luminosity in units of $10^{40} \ergsec$,\\
0.5 - 8 keV.
\end{tabular}
\label{lumdiags}
\end{table}

We can attempt to place the X-ray characteristics of NGC 4485/90
derived from our \chan observation in context by comparison with \chan
studies of other starbursting systems.  This is unfortunately not yet
possible for the prototypical starbursts in NGC 253 and M82, where
initial \chan studies have focussed only on specific facets of their
nuclear regions (e.g. the ULX in M82, Kaaret et al. 2001; the nuclear
outflow of NGC 253, Strickland et al. 2000).  However, we can draw
useful comparisons with the vigorous starbursts in the Antennae (NGC
4038/9) and NGC 3256 (Fabbiano, Zezas \& Murray 2001; Lira et
al. 2002), on the basis of the statistical study of the links between
star formation activity and the X-ray properties of spiral galaxies
derived by Read \& Ponman (2001).  They find that in starburst
galaxies the luminosity of diffuse emission per unit mass scales
closely with star forming activity (measured by FIR luminosity per
unit mass), whereas in normal galaxies it scales with galaxy size.
Although it is a less well-defined relationship, it also appears that
the total source luminosity in starburst systems increases with
activity, though at a slower rate than the diffuse luminosity.  This
implies that, in general, more active starbursts should have larger
diffuse emission contributions to their overall luminosity.  We show a
comparison of the relative luminosity diagnostics for the three
systems in Table~\ref{lumdiags}.  Both the Antennae and NGC 3256 are
more active than NGC 4485/90 on the basis of the L$_{FIR}$/L$_B$
criterion used by Read \& Ponman (2001).  They are also considerably
more X-ray luminous, with overall luminosities of $\sim 8 \times
10^{40} \ergsec$ and $\sim 3.5 \times 10^{41} \ergsec$ (0.5 - 8 keV)
respectively\footnote{We use a $H_0$ value of 75 km s$^{-1}$
Mpc$^{-1}$ here, as opposed to a value of 50 km s$^{-1}$ Mpc$^{-1}$
adopted by both Fabbiano et al. (2001) and Lira et al. (2002), and
convert X-ray fluxes into the 0.5 - 8 keV band according to spectral
models used by the respective authors.}.  Importantly, both systems
display a far higher fraction of diffuse X-ray emission relative to
the point-like source contribution, with $\sim 40\%$ and $\la 80\%$ of
the flux originating in a diffuse component in each case.  This is, of
course, consistent with the trends identified by Read \& Ponman
(2001), with the stark contrast in X-ray luminosities and diffuse
fractions emphasizing that the Antennae, and in particular NGC 3256,
are towards the high X-ray luminosity end of the starburst activity
scale, in contrast to the relatively low activity of NGC 4485/90.

The comparison between the Antennae and NGC 4485/90 is particularly
interesting given that they have many distinct similarities, such as
the observed star formation rate, their L$_{FIR}$/L$_B$ ratios and
their identical X-ray source luminosity function slopes, but such
disparate diffuse X-ray emission properties.  Major differences exist
between the systems in the type and the stage of interaction they are
undergoing; in NGC 4485/90 we see the aftermath of a close prograde
encounter between a (comparatively) large galaxy and its smaller
companion, with the tidal disruption of both galaxies as a result,
whereas in the Antennae we are witnessing the advanced stages of a
direct merger between two similar-sized galaxies.  This may suggest
that it is not so much the star formation induced by the merger that
powers the additional diffuse X-ray emission in the Antennae, but the
merger itself, perhaps through processes such as the direct shock
heating of the ISM in the merger.  An alternative explanation for the
difference in diffuse components may be the actual size of the
galaxies; if we roughly scale the mass of each system to its blue
luminosity (as per Read \& Ponman 2001), then we expect the Antennae
to be $\sim 10$ times more massive than NGC 4485/90.  It will hence
have a deeper gravitational potential well in which to retain the hot,
diffuse gas.  We note that this particular argument is espoused by
Read \& Ponman (2001) for ``normal'' galaxies in which more massive
systems appear to have a higher diffuse gas content per unit mass.
This effect may be accentuated in the case of NGC 4485/90 where the
bulk of the current star formation appears to be occurring away from
the centre of the system, where the hot gas will find it easier to
escape the gravitational influence of the galaxies.

\section{The relationship between ULX and star formation}

In section 3.2 we detailed the characteristics of the six ULX we find
in the \chan observation of NGC 4485/90.  All are found to be
point-like at the \chan spatial resolution of 0.5$''$ ($\equiv 20$ pc
in NGC 4485/90), meaning they are likely to be individual sources and
not composite X-ray emitting regions.  The best diagnostics of their
nature come from their X-ray spectra and long-term lightcurves, which
imply that five of the ULX have properties consistent with a black
hole X-ray binary, whilst the sixth (CXOU J123029.5+413927) is
probably an X-ray luminous supernova remnant.  But why are so many ULX
found in this relatively small galaxy pair?

Recent \chan observations have revealed that very active star forming
systems appear rich in ULX, for instance with the detection of 8 ULX
in NGC 3256 and a further 12 in the Antennae (Lira et al. 2002;
Fabbiano et al. 2001).  ULX are detected in other starburst systems,
notably the very luminous ULX close to the nucleus of M82 (Kaaret et
al. 2001) and a similar near-nuclear source in NGC 3628 (Strickland et
al. 2001).  However, observations of galaxies with relatively low star
formation rates tend to find few sources with luminosities in excess
of $10^{38} \ergsec$, and none in the ULX regime, with perhaps the
best example being our nearest major neighbour M31 (e.g. Supper et
al. 2001).  Furthermore, many ULX appear to be spatially coincident
with the star formation regions in systems such as the Antennae
(Fabbiano et al. 2001).  In fact the first reported optical
counterpart to a ULX is coincident with a young ($< 10$ Myr old)
stellar cluster in NGC 5204 (Roberts et al. 2001; Goad et al. 2002).
This evidence points to a physical link between the presence of many
ULX and active star formation.

The observation of NGC 4485/90 presented in this paper strongly
supports this relationship.  First and foremost we again detect a
comparatively large number of ULX in a starbursting system.
Furthermore, all the ULX are seen in or very close to regions of
active star formation (c.f. Figure~\ref{dss2srcs}).  Four are
associated with the star formation ongoing in the tidally-disrupted
spiral arms between the galaxies, a fifth is in the star forming
regions close to the nucleus of NGC 4490, and the sixth is in the
eastern spiral arm, at the base of the tidal tail.  An additional
important result is that we are able to show that the majority (5/6)
of these ULX are probably black hole X-ray binary systems, meaning
that, even in star forming environments, X-ray luminous supernova
remnants provide only a small proportion of the ULX population.

It has been suggested that ULX may constitute a new population of
$10^2 - 10^4$ M$_{\odot}$ {\it intermediate-mass\/} black holes based
upon their X-ray luminosities exceeding the Eddington luminosity for a
10 M$_{\odot}$ black hole of $\sim 10^{39.3} \ergsec$ (e.g. Colbert \&
Mushotzky 1999).  However, the formation of such objects is
problematic.  One possible origin may be through the hierarchal
merging of lower mass black holes within a dense stellar cluster
(e.g. Taniguchi et al. 2000; Ebisuzaki et al. 2001), though this may
only be applicable in the circumnuclear regions of galaxies where the
potential well is deep enough to retain the newly-formed black holes
within the clusters.  It is also a long-timescale process, taking
$\sim 10^9$ years.  Elmegreen et al. (1998) estimate the ages of the
stellar populations within distinct regions of the NGC 4485/90 system.
The intermediate-mass black hole formation timescale is inconsistent
with the age of the main stellar populations in both the spiral arms
of NGC 4490 ($\sim 4 \times 10^8$ years) and the tidal tail of NGC
4485 ($\sim 10^8$ years), where 4/5 of the putative black hole X-ray
binary ULX are located, though it is closer to the age of the bulge of
NGC 4490 ($\sim 6 \times 10^8$ years).  A timescale of $\sim 10^9$
years is of course also inconsistent with the ULX being directly
related to the ongoing star formation.  We can therefore rule out this
formation scenario for 4/5 of the black hole X-ray binary ULX, though
we note that the near-nuclear ULX CXOU J123036.3+413837 may be a
candidate intermediate-mass black hole formed in a dense cluster based
on its position and the local stellar age; alternatively its presence
may simply be related to the recent star formation in the nucleus.

A second method of forming intermediate-mass black holes is through
the collapse of primordial Population III objects (Madau \& Rees
2001).  This would have left a numerous population of
intermediate-mass black holes, many of which will have migrated to the
centres of galaxies through the action of dynamical friction (and in
many cases merged with the supermassive black hole).  However, a
number may still be present away from the galaxy nuclei.  This origin
for the ULX in NGC 4485/90 is also very unlikely, given that the ULX
only appear coincident with the regions of recent star formation.

It therefore appears unlikely that at least four, if not all five, of
the putative black hole X-ray binary ULX in NGC 4485/90 are
intermediate-mass black hole systems.  This leaves the most likely
remaining alternative that they are ordinary stellar mass X-ray binary
systems that are apparently radiating in excess of the Eddington
luminosity.  This is readily explained if the X-ray emission of the
sources is anisotropic. King et al. (2001) suggest that this is indeed
the case, with the X-ray emission of the ULX mildly beamed into our
line-of-sight.  Beaming, however, requires large numbers of potential
sources because at any one time many more of the sources must exist
than are seen along a single line-of-sight.  King et al. (2001)
resolve this problem by suggesting that ULX are ``ordinary'' black
hole and/or neutron star high- and intermediate-mass X-ray binaries
undergoing a thermal-timescale mass transfer epoch that is an
inevitable stage of binary evolution.  This in turn suggests a link to
Galactic microquasars.  The association of ULX with intense star
formation activity is then readily explained by the relatively short
lifetimes of the high-mass X-ray binary systems formed in large
numbers in such an epoch.

The possible link between ULX and Galactic microquasars suggested by
King et al. (2001) is also supported on the basis of \asca
spectroscopy (Makishima et al. 2000).  Georganopoulos, Aharonian \&
Kirk (2002) expand upon this idea, advocating that ULX are consistent
with microquasars with high stellar mass companions.  In their model
ULX are microquasars seen with the jet oriented close to our
line-of-sight, and the low-hard state emission originates in Compton
scattering of photons from the companion star or accretion disk by
relativistic electrons in the jet.  Further support for this model
comes from K{\"o}rding, Falcke \& Markoff (2002), who find that a
simple population synthesis model for beamed X-ray sources can
adequately explain the observed populations of luminous X-ray sources
in nearby galaxies.

Other solutions to the problem of producing apparently super-Eddington
accretion onto a stellar-mass compact object are discussed by Grimm,
Gilfanov \& Sunyaev (2002).  Most interestingly, they note that a
combination of a face-on viewing aspect for the accretion disk and an
unusual chemical composition of the secondary star (e.g. He-enriched)
can combine to apparently exceed the Eddington luminosity by a factor
$\sim 6$.  Given that the intrinsic luminosities of the NGC 4485/90
ULX lie only a factor $\sim 1.5 - 2.5$ above the Eddington luminosity
for a 10 M$_{\odot}$ black hole, then this is a very plausible
explanation for their super-Eddington luminosities.  

A final possible alternative to intermediate-mass black holes is that
the accretion is literally a super-Eddington process.  This has
recently been proposed by Begelman (2002), who shows that a ``thin''
accretion disk in a radiation-pressure dominated (i.e. high accretion
rate) regime can produce fluxes up to a factor 10 above the Eddington
limit for a stellar-mass black hole X-ray binary.  Similarly, the
``slim disk'' scenario (e.g. Abramowicz et al. 1988) is also predicted
to produce super-Eddington fluxes that could account for ULX (Watarai,
Mizuno \& Mineshige 2001).  The high inner-accretion disk temperatures
observed by Makishima et al. (2000) and noted in this paper for the
ULX in NGC 4485/90 are consistent with the stellar mass black holes
inferred in both the above models.

\section{Conclusions}

In this paper we have analysed a 20 ks \chan observation of the
nearby, interacting galaxy pair NGC 4485/90.  Our results may be
summarised thus:

\begin{itemize}
\item
We detect a total of 31 discrete X-ray sources coincident with the
galaxies (including a faint bridge linking the two), ranging in
observed luminosity between $\sim 2 \times 10^{37}$ and $4 \times
10^{39} \ergsec$ (0.5 - 8 keV).  Only five had been detected in
previous X-ray observations of the galaxies (RPS97; RW2000).  No more
than four of these detections are expected to be due to background
contamination.
\item
An analysis of the 0.3 -- 2 keV vs 2 -- 10 keV hardness ratios for the
sources reveals a trend of spectral softening as the flux decreases.
This is not a result of a new, soft source population appearing at
lower fluxes (though we do identifying one candidate super-soft
source), but rather the fact that the more luminous sources are more
absorbed.  This absorption may be intrinsic to the sources, or may be
a result of their being located deeper within, or in denser regions
of, the galaxies.
\item
We identify six sources with luminosities in excess of $10^{39}
\ergsec$, which are examples of the ultraluminous X-ray source
phenomenon.  All of these sources are point-like and none exhibit
short-term X-ray variability.  On the basis of their X-ray spectra and
long-term lightcurves five of the ULX are likely to be black hole
X-ray binary systems and the sixth, which is directly coincident with
a radio source, is an X-ray luminous supernova remnant.
\item
There is extensive diffuse X-ray emission associated with the disk of
NGC 4490.  A further patch is located in the southern regions of NGC
4485, probably as a result of the ram-pressure stripping of its ISM as
it passes through the giant \hi cloud surrounding NGC 4490.  X-ray
spectroscopy of the diffuse component shows it to be predominantly
thermal in nature, consistent with an origin in the hot interstellar
medium of the galaxies, and not in an unresolved source population.
\item
The total observed X-ray luminosity of the NGC 4485/90 system is $\sim
2 \times 10^{40} \ergsec$ (0.5 - 8 keV).  Over 90\% of this is
contributed by the luminous X-ray source population, with the three
brightest sources contributing more than 50\% between them.  The
luminosity function of the discrete X-ray source population shows a
very flat slope ($-0.57 \pm 0.1$) similar to that seen in classical
starburst systems, implying that NGC 4485/90 hosts a young, luminous
X-ray source population.
\item
A comparison with the classic galaxy merger starbursts in the Antennae
and NGC 3256 shows that the starburst in NGC 4485/90 is relatively
X-ray weak, particularly in terms of its diffuse emission component.
This is especially puzzling when compared to the Antennae, which has a
similar activity level.  This may be due to the type of interaction
(direct merger versus tidal encounter), or the smaller gravitational
potential well of NGC 4490 retaining less hot gas.
\item
The high ULX content of NGC 4485/90, and the direct spatial
coincidence of its ULX with star formation regions, is further
convincing evidence of the relationship of many ULX with active star
formation.  We note that only one of the ULX is identifiable as a
supernova remnant, implying that even in active star formation regions
the ULX population is dominated by accreting sources.
\item
It is unlikely that the putative black hole X-ray binary ULX in NGC
4485/90 contain intermediate-mass black holes, due to their
extra-nuclear positions, the age of the stellar populations in their
host regions, and their apparent relationship with the active star
formation regions within the galaxies.  There are a number of very
plausible alternative explanations in terms of a simple high-mass
X-ray binary nature for the ULX.  It now seems credible that many ULX
may be nothing more exotic than ordinary high-mass X-ray binary
systems viewed in an epoch of unusually high accretion rate and/or
with a preferred orientation.
\end{itemize}

The most remarkable feature of NGC 4485/90 in the X-ray regime is its
highly luminous discrete X-ray source population, and in particular
the six ULX it hosts.  Indeed, NGC 4485/90 may present one of the best
opportunities to study this phenomenon in the local universe, since
other systems with similar numbers of ULX (e.g. the Antennae and NGC
3256) are over a factor 2.5 more distant, and dominated by
potentially-confusing extensive hot gas components.  A better
understanding of the phenomenology of this interesting class of
sources may therefore rely upon future studies of the ULX in NGC
4485/90.

\vspace{0.2cm}

{\noindent \bf ACKNOWLEDGMENTS}

We thank the referee, Andy Read, for his many useful comments that
have improved this paper.  TPR gratefully acknowledges financial
support from PPARC.  This paper uses DSS-2 data extracted from the ESO
online digitised sky survey.  The archival {\it ROSAT\/} data were
obtained from the Leicester database and archive service (LEDAS) at
the University of Leicester.

\end{document}